\documentclass[preprint]{aastex}

\newcommand{\hi}{\ion{H}{1}}
\newcommand{\etal}{{\it et al.} }
\newcommand{\kms}{km s$^{-1}$}

\slugcomment{\hi\ in KISS Dwarfs - 08Aug02}

\shorttitle{\hi\ in KISS Dwarf Galaxies}
\shortauthors{Lee et al.}

\begin{document}


\title{\hi\ Properties of Low Luminosity Star-Forming Galaxies\\ in the KPNO International Spectroscopic Survey}


\author{Janice C. Lee \altaffilmark{1}}
\affil{Steward Observatory, University of Arizona, Tucson, AZ 85721}
\email{jlee@as.arizona.edu}
\altaffiltext{1}{Visiting Scientist, Arecibo Observatory.  The Arecibo radio telescope is part of the National Astronomy and Ionosphere Center and is operated by Cornell University under a management agreement with the National Science Foundation.}

\author{John J. Salzer \altaffilmark{1}}
\affil{Astronomy Department, Wesleyan University, Middletown, CT 06457}
\email{slaz@astro.wesleyan.edu}

\author{Chris Impey \altaffilmark{1}}
\affil{Steward Observatory, University of Arizona, Tucson, AZ 85721}
\email{cimpey@as.arizona.edu}

\author{Trinh X. Thuan \altaffilmark{1}}
\affil{Astronomy Department, University of Virginia, Charlottesville, VA}
\email{txt@virginia.edu}

\and

\author{Caryl Gronwall \altaffilmark{1}}
\affil{Department of Physics \& Astronomy, Johns Hopkins University, Baltimore, MD 21218}
\email{caryl@adcam.pha.jhu.edu}



\begin{abstract}
New HI observations are presented for a complete sample of 109 low luminosity star-forming galaxies taken from the KPNO International Spectroscopic Survey (KISS), the first CCD-based wide-field objective-prism survey for emission-line galaxies.  This sample consists of all star-forming galaxies with $M_B > -18.0$ and $cz <$ 11,000 km s$^{-1}$ from the first H$\alpha$-selected survey list.  The galaxies in this list lie within a 1.3 deg wide strip centered on $\delta($B$1950) = 29\arcdeg30\arcmin$ that spans the range $\alpha($B$1950)=12^{\mbox{\scriptsize h}}15^{\mbox{\scriptsize m}}$ to $\alpha($B$1950)=17^{\mbox{\scriptsize h}}0^{\mbox{\scriptsize m}}$.  Overall, 97 out of 109 galaxies have been detected in HI.  We confirm the weak trend of increasing gas richness with decreasing luminosity found by previous authors.  Gas richness is also shown to be weakly anti-correlated with metallicity.  The dependence of star formation rates (SFRs) and HI gas depletion timescales on metallicity is examined.  The median solar metallicity based SFR and gas depletion timescale are 0.1639 M$_{\odot}$ yr$^{-1}$ and 5 Gyrs, respectively.  Corrections for variations in metallicity decreases SFRs by $\sim0.5$ dex and increases gas depletion timescales by an average of $\sim$8 Gyrs.  The majority of galaxies in this sample still have large reservoirs of HI gas, and despite their large current star formation rates, could have formed stars in a quasi-continuous manner for a Hubble time.  Finally, we present the first HI mass function for low luminosity star-forming galaxies and show that this subpopulation contributes 10-15\% of the overall HI density in the local universe.  We conclude that if the HI mass function of the Universe does indeed have a steeply rising low-mass slope as suggested by previous authors, it is not due to the population of low luminosity star-forming galaxies.  Comparison of the number densities from the HIMFs in the range $10^8<M_{HI}/M_{\odot}<10^9$ implies that  25-50\% of galaxies in this regime are currently undergoing a strong episode of star formation.

\end{abstract}


\keywords{galaxies: dwarf ---  galaxies: irregular --- galaxies: ISM --- galaxies: luminosity function, mass function --- galaxies: starburst --- ISM: HI}


\section{Introduction}

In an attempt to develop a unified understanding of the different morphological classes of dwarf galaxies, efforts have been made to relate the various observed types through some evolutionary scheme.  A number of scenarios have been proposed (see Skillman \& Bender 1995, Ferrara \& Tolstoy 2000 for excellent reviews).  One possibility is that the gas-rich dwarf irregulars (dIs) go through an extremely strong burst of star formation, lose the majority of their gas after the starburst event, and eventually fade and evolve into the gas-poor dwarf ellipticals (dEs).  During the bursting phase, the dIs would resemble massive HII regions.  Another possibility is that the dIs and dEs have common progenitors but have evolved differently because of environmental effects.  Although these schemes are highly debated, they form a good context for investigating the comparative structural properties of the different classes of dwarfs.

Several questions arise when considering these evolutionary scenarios.  Can {\it any} dI initiate a large star-formation episode and appear as a BCD?  In other words, is the BCD phenomenon a stage of galaxy evolution common to all gas-rich dwarfs or only to a select sub-sample with special characteristics?  Further, do the progenitor galaxies go through repeated intense bursts of activity, or is the bursting phase unique in the object's lifetime?  If the bursts repeat, then what fraction of the time is the galaxy bursting (i.e., what is the duty cycle)?  Answers to these questions can also help us to determine if starbursting dwarfs can be used as a fair tracer of the more general population of low-mass galaxies.  Since starbursting dwarfs are far more easily detected than their quiescent counterparts, this issue is of great importance to investigations of the faint-end/low-mass tails of the luminosity/HI mass functions, and to studies of large-scale structure, particularly since low-mass galaxies may be less biased and better tracers of the overall mass distribution (e.g. Santiago \& da Costa 1990, Loveday et al. 1995, Lee et al. 1999).

Within this context, it is interesting to study the HI properties of low luminosity star-forming galaxies, since the quantity of HI indicates the amount of fuel available for driving a starburst.  Additional constraints may be placed on the star-formation history  and chemical evolution of bursting dwarfs when HI data are used in conjunction with the star-formation rates and abundances inferred from optical spectroscopy.  Also, dynamical masses of these systems may be estimated from the velocity widths of HI profiles.  With this combination of optical and radio information, it is possible to assess the plausibility of gas-rich dIs as progenitors of bursting dwarfs, and to attempt to answer the questions posed above.   One can further study the differences in the distribution of HI in various morphological types by obtaining HI interferometric maps.  This is generally only done after initial single-dish HI observations have been made and viable candidates for interferometric mapping have been identified.  Hence, observations of the neutral gas content of bursting dwarfs is an important step toward a more complete understanding of this class of galaxy.

Previous 21-cm observations of star-forming dwarf galaxies (Thuan \& Martin 1981; Gordon \& Gottesman 1981; Staveley-Smith, Davies \& Kinman 1992; Thuan et al. 1999; Smoker et al. 2000; Salzer et al. 2002) have been fruitful in establishing the properties of their neutral gas.  However, the samples used in these previous studies are comprised of objects from a number of different catalogs.  For example, Thuan and Martin (1981) used a combination of dwarfs from the Markarian, Haro and Zwicky catalogs, while Salzer et al. (2002) selected dwarfs from the Michigan, Wasilewski and Case surveys, in addition to the catalogs used by Thuan and Martin.  HI observations are needed for a complete, uniformly selected sample of dwarf galaxies since the studies which are based on such heterogeneous collections of objects may be biased in unrecognized ways.  

To remedy this situation, we have obtained HI observations for a new, ultra-deep, complete sample of actively star-forming, low luminosity galaxies using the Arecibo Observatory 305-m radio telescope.  Our sample consists of low luminosity (M$_{B} \geq -18.0$), nearby ($cz \leq $11,000 km s$^{-1}$) , H$\alpha$ selected galaxies from the first survey strip of the KPNO International Spectroscopic Survey (KISS) (Salzer et al. 2000, hereafter Paper I; Salzer et al. 2001, hereafter KISS Red 1 [KR1]), a new objective prism survey for extragalactic emission-line objects.  In this paper, we present the 21-cm HI line observations of the overall sample.  The remaining sections are organized as follows.  An overview of the sample and observational procedures are given in $\S$ 2.  The HI data are presented $\S$ 3, and the HI properties are described in $\S$ 4.  Finally, our conclusions are summarized in $\S$ 5.  Suggestions for further work involving the dwarf candidates in this sample are given throughout the paper.  All distance dependent quantities assume H$_{\mbox{\scriptsize o}} = $75 km s$^{-1}$ Mpc$^{-1}$.
\section{\hi\ Observations}

\subsection{Sample Overview}

Objective-prism surveys which select objects on the basis of line-emission, like the KPNO International Spectroscopic Survey (KISS, see Paper I \& KR1), are excellent sources of low luminosity galaxies.  Because the equivalent-widths of the targeted lines generally become larger as the luminosity of the host galaxy decreases, this technique is extremely effective at detecting intrinsically faint objects and partially helps to overcome the constraints imposed by the Malmquist effect on surveys which are magnitude-limited (see Figure 8 in KR1).  In particular, since KISS is the first wide-field objective-prism survey to take advantage of the capabilities of CCD imaging, it has yielded a sample of emission-line galaxies (ELGs) which is at least two magnitudes fainter than previous line-selected Schmidt surveys, providing us with one of the deepest available samples of dwarfs (see Table 2 in Paper I).  

Our sample of low luminosity star-forming galaxies is taken from List 1 of the KISS catalog (KR1).  The portion of sky covered by List 1 was chosen to overlap the Century redshift survey (Geller et al. 1997, Wegner et al. 2001), so that comparison with a deep magnitude-selected survey would be possible.  This area is a 1.3 deg wide strip centered on $\delta($B$1950) = 29\arcdeg30\arcmin$ spanning the range $\alpha($B$1950)=12^{\mbox{\scriptsize h}}15^{\mbox{\scriptsize m}}$ to $\alpha($B$1950)=17^{\mbox{\scriptsize h}}0^{\mbox{\scriptsize m}}$.  The ELGs are H$\alpha$ selected and have been cataloged in a uniform way (Paper I).  The overall sample from List 1 has a well-understood selection function which is based on the line plus continuum flux at H$\alpha$ (Gronwall et al. 2002, also see discussion in $\S$4.5).

We initially specified selection criteria that required targets to have: (1) spectra characteristic of star-forming galaxies (HII-region-like spectra which exhibit strong narrow emission lines superimposed on a stellar continuum which rises towards the blue), (2) $cz < 11,000$ \kms, and (3) $M_B > -18.0$.  When HI observations for this sample began in May 2000, however, follow-up slit spectral observations had not been completed for the entire List 1 catalog.  At that time target selection was based on the objective-prism estimates of the redshifts for those galaxies that lacked slit spectra.  Recessional velocities derived from the coarse (24 \AA/pixel) survey spectra have a 1$\sigma$ error of $\pm$830 \kms.  About 40\% of the galaxies in our original HI sample did not have follow-up spectra.  

Slit spectral data was subsequently obtained at various telescopes for the remainder of the HI sample over the following spring observing season (Melbourne et al. 2002; Wegner et al. 2002; Salzer et al. 2003).  When redshifts and blue absolute magnitudes were recalculated from the new data, it was found that 9 of our original targets no longer met our luminosity and/or recessional velocity criteria.  However, three of these galaxies had absolute magnitudes only nominally brighter than $-18.0$, so we decided to retain these in the sample (see Table 1 and Figure 1).  An additional 5 galaxies which were not in our original sample were then found to be within the specified limits.  These were added to the sample towards the end of our 2001 observing run. 

The final HI sample contains a total of 109 galaxies.  When compared with the overall KISS sample, the HI sub-sample has a brighter median $m_B$ (17.54 versus 18.08), which indicates that the low-luminosity galaxies are generally found nearby.  This is due in large part to the velocity limit of 11,000 \kms imposed on the HI sample.  The median velocity of the HI sample (7809 \kms) is only $\sim$40\% the median velocity of the full KISS sample (18,407 \kms).  For comparison,  the median m$_B$ of the Salzer et al. (2002) HI sample of dwarf star-forming galaxies is 15.7, while that of galaxies observed in the Staveley-Smith et al.(1992) study is 15.1.  In other words, our current sample is substantially fainter than those found in all previous studies of this class of galaxy.

The median $B-V$ color of the KISS HI sample, 0.51, is much bluer than that of the parent sample, which has a median $B-V$ of 0.67, the typical color of an Sb galaxy (Roberts \& Haynes 1994).  It is interesting to note that the value of 0.51 is comparable to the median $B-V$ of 0.54 for the [OIII]-selected University of Michigan objective-prism survey (Salzer et al. 1989).  This is one indication that the low-luminosity, low-metallicity galaxies preferentially found in [OIII] surveys are also included in H$\alpha$ selected surveys (see Paper I, $\S$4 for further discussion).  The luminosity distribution of the HI sample (Figure 1) has a median of $-$16.87, which is about the luminosity of the Small Magellanic Cloud (van den Bergh 2000).  Overall, our targets are brighter that those found in HI surveys of dwarf galaxies such as that of Salzer et al. (2002) which has median $M_B$ of $-$16.1, and Staveley-Smith et al. (1992) who imposed a luminosity limit of $M_B= -$16.0.

\subsection{Observational Procedures}

HI observations of 109 local ($cz < 11,000$ \kms), low luminosity ($M_B > -18.0$) KISS emission-line galaxies were carried out with the Arecibo Observatory 305-m radio telescope during May 2000 and from May to July 2001.  Approximately 90\% of our data was taken with the L-narrow receiver at the upgraded Gregorian feed.  This receiver functions at an average system temperature of 30 K.  The less sensitive L-wide receiver (T$_{\mbox{\scriptsize sys}} \sim$ 38 K) was used for three nights in May 2000, when the L-narrow receiver was off-line.  The four subcorrelator boards were each configured with 2048 channels and centered at the recessional velocity found from the target's optical spectrum.  Nine-level sampling was used.  Two boards were operated in a ``lower-resolution mode'' using a 25 MHz bandwidth, resulting in a channel spacing of 2.6 \kms at 1420 MHz, while two boards were used in a ``higher-resolution mode'' using a 12.5 MHz bandwidth, resulting in a channel spacing of 1.3 \kms at 1420 MHz.  The boards in each of the resolution modes recorded signals in independent, opposite circular polarizations, which are subsequently combined in the reduction process.

Observations were made in total power mode with 5 minute on-source, 5 minute off-source pairs, followed by 10 second on-off noise diode calibration pairs.  During the integrations, data dumps occurred every 6 seconds.  This high sampling rate allowed for spurious, intermittent radio frequency interference (RFI) to be excised during the reduction process without sacrificing large amounts of integration time.  With the exception of a few particularly strong lined sources, a minimum of 4 on-off pairs were taken for each  target.

When possible, observations were made after astronomical sunset to avoid problems with solar interference.  However, since many of our time blocks were scheduled to begin around sunset, about 10\% of our observations show its effects and have sinusoidal variations in their baselines. (For example, see KISSR 1 in the first panel of Figure 2.)  Notes on other galaxies similarly affected are given in Table 2.  System checks were performed at the beginning of each run by observing strong, non-variable continuum point sources from an unpublished list of unconfused calibrator sources prepared from the NRAO VLA Sky Survey (NVSS) by Jim Condon and Qi Feng Yi, which is made available to observers through a catalog in the Arecibo telescope control GUI. 

\subsection{Data Reduction}

The data were reduced with the ANALYZ software package, using appropriate L-band gain, zenith angle and azimuth corrections for the refurbished Arecibo dish and standard averaging and calibration routines.  Intermittent broadband interference, such as the GPS L3 signal at 1381 MHz, were removed by entirely omitting the 6-second scans in which they appeared from the averaging procedure.  Narrow single-channel interference spikes were manually removed and the average value of the two adjacent channels was substituted.  In both cases, RFI was removed only when it appeared within 500 km s$^{-1}$ of the center of the observed line.  Baseline fitting, removal and line measurements were performed with the GALPAC analysis system developed by R. Giovanelli.

\section{HI Data}

\subsection{Observational Results}
Figure 2 presents the 25 MHz bandwidth HI profiles for all 109 galaxies in the sample.  Five-channel boxcar smoothing has been applied and baseline polynomial fits to the data are over-plotted.  In addition, $4\farcm5 \times 4\farcm0$ $B$ and $V$-band composite KISS survey images of the targets are shown in Figure 3.  Identification numbers as assigned in KR1 are given in the upper left hand corner of the images, while the KISS field number and field object ID are given in the upper right hand corner.  Observed quantities are reported in Table 1, and the column entries shown there are as follows:
\renewcommand{\labelenumi}{(\arabic{enumi})}
\begin{enumerate}

\item Galaxy identification number as given in the KISS Red survey strip (KR1).  Objects for which notes are given in Table 2 are marked by an asterisk. 

\item Right Ascension in J2000 coordinates.

\item Declination in J2000 coordinates.

\item KISS $B$-band apparent magnitude, from KR1.

\item KISS $B$-Band absolute magnitude.  Distances are calculated using relativistically correct recessional velocities which are adjusted for the rotation of the Galaxy and its motion about the mass centroid of the Local Group.  The standard relation  $cz$ = $cz_{\odot}$ + 300 km s$^{-1}$ sin $l$ cos $b$ is used and, H$_{\mbox{\scriptsize o}} =$ 75 km s$^{-1}$ Mpc$^{-1}$ is assumed.

\item KISS $B - V$ color, from KR1.

\item Heliocentric $cz$, v$_{helio}$, measured at the midpoint of the 21-cm line profile at 50\% of the peak flux.  For non-detections, $cz$ measured from the object's optical spectrum is given in parentheses.

\item Velocity width, $\Delta$v$_{50}$, at 50\% of the peak flux in km s$^{-1}$.  No data are listed for non-detections.  Widths are given in parentheses for 7 of the confused sources, while no data are listed for the 3 most severely confused sources (see $\S$ 3.2 \& 3.3).

\item Velocity width, $\Delta$v$_{20}$, at 20\% of the peak flux in km s$^{-1}$.  No data are listed for non-detections.  Widths are given in parentheses for 7 of the confused sources, while no data are listed for the 3 most severely confused sources (see $\S$ 3.2 \& 3.3).

\item The observed integrated 21-cm HI line flux, F.I. $= \int S(v)dv$, in Jy km s$^{-1}$.  Upper limits, discussed in Section 3.3, are listed for the 12 non-detections in the sample.  Upper limits or adjusted fluxes are shown for 7 of the confused sources, while no data are listed for the 3 most severely confused sources (see $\S$ 3.2 \& 3.3).  Corrections for beam size are only necessary for 14 objects, and the observed fluxes of these galaxies are given in parentheses.  The calculation of the correction factors is discussed in Section 3.4 and the factors themselves are listed in Table 3.

\item RMS deviation in the baseline fit to the 5-channel boxcar smoothed 21-cm spectrum, in mJy.  This quantity is used to compute upper limits for the undetected sources (see $\S$ 3.3).

\item Order of the polynomial used in the baseline fit.

\item Total on-source integration time after scans containing RFI near the velocity of the target profile have been removed.

\item Signal-to-Noise ratio, $SNR$ = F.I./($\sqrt{N_{ch}}\cdot RMS\cdot\Delta v_{res}$), where the velocity width at 20\% of the mean flux divided by $\Delta v_{res}$, the channel spacing prior to smoothing, is adopted for the $N_{ch}$, the number of channels over which the line is detected.  For all entries in the table, $\Delta v_{res}$ = 2.6 km s$^{-1}$.

\end{enumerate}

Overall, 97 out of 109 galaxies were detected in HI, resulting in a detection rate of $\sim$89\%.  Of these 97, eight are weak detections with 2.5 $< SNR <$ 5.8.  Our non-detections are generally objects that were added to the sample towards the end of our Arecibo observing run in 2001 (see $\S$2.1) and that suffered from solar interference and/or insufficient integration times.  They are not preferentially galaxies that have weak 21-cm signals. 

\subsection{Confused Sources}

It is important to pay careful attention to the possibility of source confusion within the Arecibo beam for this particular sample.  This is because the majority of targets are very small compared to the $\sim 3\farcm5$ half power beam width, and the linear distance subtended by the beam at the average recessional velocity for our sample ($\sim$8000 \kms) is 0.1 Mpc, which is on the order of the mean projected separation of a loose association of dwarf galaxies.  

To check for possible contamination of the observed fluxes by nearby companions, the area within a radius of 7\arcmin\ of each of the targets was searched for potential neighbors.  This examination, which was performed on the KISS composite $B$ and $V$-band survey images, covers an area which includes both the telescope's main beam and first sidelobe, whose peak is located $\sim 5\farcm5$ away from the peak of the main lobe and has a relative intensity of $\sim$5\% (Heiles et al. 2001).  Our determination of actual confusion was based on three factors: (1) the transverse distance of the neighbor to the target, (2) the optical brightness of the neighbor compared with the target, and (3) the velocity difference between the neighbor and the target.  The KISS catalog, Century Redshift Survey catalog and the NASA Extragalactic Database (NED) were all searched for redshifts of the potential neighbors. 

After the first round of examinations, it was concluded that neighboring objects which appeared within the sidelobes, but outside of the main beam, did not significantly contribute to the observed flux of the target.  This conclusion was based on examples of bright galaxies that appeared within the sidelobes and were offset from the target velocity by several hundred km s$^{-1}$, but did not produce observable emission at the velocity reported in the literature.  The clearest illustrations of this are in the spectra of KISSR 96 and KISSR 97.  UGC 7836, an edge-on Scd galaxy with $m_B$ = 14.83, is located 5\farcm5 to the NE of KISSR 96 and 5\farcm6 to the SE of KISSR 97 at 9337 km s$^{-1}$.  As anticipated, this galaxy has a strong, broad ($w_{50} \sim$ 480 km s$^{-1}$) double-horned profile (Haynes et al. 1997), characteristic of a gas-rich, late-type spiral.  However, in the spectrum of KISSR 96, there is only one narrow line centered on 9200 km s$^{-1}$ with no emission appearing above $\sim$9300 km s$^{-1}$.  Meanwhile, in spectrum of KISSR 97, although there {\it is} a double-horned profile, it is centered on the optical velocity of the target.  Further, the HI profile of KISS 97 is fairly narrow ($w_{50} \sim$ 190 km s$^{-1}$) compared to the width of U7836's 21-cm line, and no emission appears below $\sim$ 9300 km s$^{-1}$.  CGCG 159-053, an interacting galaxy pair which is 5\farcm4 E of KISSR 97 at 6905 km s$^{-1}$, is not detected either.  Therefore, since the sidelobe companions of all of the other targets in our sample are substantially fainter than U7836 or CGCG 159-053, they were not considered to be sources of contamination.

Thus, observations are considered to be confused only if the companions (1) are within the main beam, (2) have optical brightnesses greater than or comparable to the target and (3) have published redshifts that place them within 200 km s$^{-1}$ of the target.  In this sample, there are 10 galaxies that meet these criteria.  They are KISSR 146, 148, 256, 257, 265, 356, 401, 404, 405, and 1013.  For five of these galaxies (256, 257, 265, 404, 405) we list conservative upper-limits for the flux in Table 1, while for two of the galaxies (356, 1013) we list fluxes which have been revised based on additional available information (see $\S$ 3.3).  No data are listed in Table 1 for three targets (146, 148, 401) that are confused with sources much brighter than the targets themselves.  Instead, the results of observing these confused regions are given in our notes on individual galaxies in Table 2.  In Table 2, observations are cited as possibly, but not likely to be confused if the target has companions which are within or on the outer periphery of the main beam, but are dimmer than the target galaxy and have no published redshift.  Twenty-four galaxies have been noted as possibly, but unlikely to be, confused.  The fluxes reported in Table 1 for these galaxies are the original observed quantities.  Alternate names for our targets given by other catalogs are also listed in Table 2.

\subsection{Upper-Limits and Revised Fluxes for Confused and Undetected Sources}

Upper limits on the fluxes of the 12 targets which were undetected in HI were calculated by assuming rectangular profiles with heights corresponding to 3$\sigma$ fluctuations in the baseline and widths based on the boundary of points in the log(M$_B$)-log(w$_{50}$) plane (Figure 4).  In Figure 4, the dashed line indicates the upper envelope of observed widths in this sample.  An undetected target's luminosity is used to find its upper-limit width using these lines.   

For 5 of the 10 confused sources, the observed composite HI flux was taken as the upper-limit.  In the spectra of KISSR 265, there is contamination from KISSR 266 ($cz$ = 9495 km s$^{-1}$), which is not in the HI sample, and from another previously unobserved galaxy 1\farcm7 SW from the target.  All of these galaxies are comparable in brightness, and the observed flux itself is taken as an upper limit.  KISSR 404 and 405 are separated by 0\farcm4 and are of comparable optical brightness and diameter.  Two independent pointings at each of these targets produce lines that are similar in strength and shape.  So, it is possible that one of the sources is responsible for all of the flux.  Again, we assign the observed flux as the upper limit for both KISSR 404 and 405.  The same reasoning holds for KISSR 256 and 257, which are separated by 1\farcm9.  

No data are listed in Table 1 for three targets that are confused with neighboring sources which are much brighter than the targets themselves (KISSR 146, 148, 401).  In the spectra of KISSR 146 and 148, the measured flux is a combination of emission from three galaxies, KISSR 146, KISSR 148 and UGC 8033 (=KISSR 147), a large spiral galaxy which is most likely producing the majority of the emission.  The HI profile of KISSR 401 is clearly a superposition of multiple lines and is due to the target itself plus two other KISS galaxies not in this sample, KISSR 399 and 400.  The target is the faintest optical source among the trio.  Observed fluxes and widths from the confused regions are instead given in the notes in Table 2.

For the remaining two confused sources, the observed fluxes are revised based on information available in the optical images and in the shape of the 21-cm line.  In the field of KISSR 356, there is a neighboring high-inclination spiral galaxy of comparable optical brightness within the beam.  Although this spiral galaxy does not have a published redshift, there are several other galaxies outside the main beam which have observed velocities that cluster around 10,500 km s$^{-1}$.  Thus, there is a strong possibility that the unidentified spiral companion is a member of this group and has a similar velocity.  Furthermore, the observed profile of KISSR 356 appears to be a superposition of a narrow-peaked source and a broad-lined source.  Thus, we assume that the broad portion of the profile originates from the spiral companion and that the flux due to the target is contained within the narrow portion of profile.  The flux and widths reported for KISSR 356 are measured by fitting a baseline to the top of the broad portion of the profile, and integrating over the narrow peak.  For KISSR 1013, arguments for the revised flux are in the inverse sense of those just given for KISSR 356.  In this case the target is a high-inclination disk-like system while the neighbor is LSB dwarf without a published redshift, and the radio spectrum again appears to be a superposition of broad and narrow emission-lines.  The reported flux and widths for KISSR 1013 is based on a measurement of the area within the broad portion of the profile.

\subsection{Beam Corrections}

Corrections to the observed flux for beamwidth are not necessary for the majority of galaxies in the sample since the apparent sizes of the galaxies are considerably smaller than the Arecibo beam.  Galaxies for which the correction factor has been determined to be greater than 3\% are listed in Table 3.  These factors have been calculated following the methods of Staveley-Smith et al. (1992) and Thuan \& Martin (1981), where the HI distribution is modeled as an elliptical Gaussian.  The correction for the source to beam size ratio is given by:
\begin{equation}
f_c = \Big[1+ \Big(\frac{a_{HI}}{\theta}\Big)^2\Big] \Big[1+ \Big(\frac{b_{HI}}{\theta}\Big)^2 \Big]
\end{equation}
where $\theta$ is the half-power beamwidth (HPBW) of the telescope, $a_{HI}$ and $b_{HI}$ are the half-power major and minor axes diameters of the neutral gas distribution, and all quantities are expressed in units of arcminutes.  The beam shape of the Arecibo telescope is elliptical with major and minor axes aligned with azimuth and zenith angle directions.  At $za = 11\degr$, the HPBW$_{za}$ is a minimum at 3\farcm66 and the HPBW$_{az}$ is a maximum at 3\farcm14, so $\theta$ is taken as 3\farcm3 (Heiles 2001).  The HI distribution is assumed to follow that of the optical light, and the HI diameters are taken to be twice the optical diameters measured at a 25 $B$-mag arcsec$^{-2}$ ellipse.  This value for $\Big(\frac{a_{HI}}{a_{25}}\Big)$ and $\Big(\frac{b_{HI}}{b_{25}}\Big)$ is based on the HI interferometric mapping of blue compact dwarfs by van Zee et al. (1998, 2001).

With the exception of KISSR 1048 (UGC 10445), the beam correction factors in Table 3 are all less than 33\% ($f_c$=1.33 for KISSR 73, a nearby diffuse low surface brightness galaxy), with $<f_c>=1.10$).  For KISSR 1048, we have computed a factor of 2.07, which yields a corrected flux integral of 36.71 Jy km s$^{-1}$.  In comparison, Haynes et al. (1998) have also observed this galaxy with the Greenbank 43-m telescope, and report an observed flux of 29.68 Jy km s$^{-1}$ with a corrected flux of 30.01 Jy km s$^{-1}$.  Given the negligible correction factor associated with the much larger beam of the 43-m telescope (HPBW = 21$\arcmin$), it would seem that the true flux of KISSR 1048 is closer to the Haynes et al. value of 30.01 Jy km s$^{-1}$ than our value of 36.71 Jy km s$^{-1}$.  The discrepancy is likely due to our adopted correction factor, which is based on a HI distribution model appropriate for starbursting dwarfs, but not for late-type spiral galaxies.  KISSR 1048 is one of the few galaxies in this sample which can be unequivocally classified as a late-type spiral, as it is face-on and nearby (v$_\odot$= 965 km s$^{-1}$).  For this class of galaxy, Hewitt et al. (1983) have established that a model with a central HI depression described by a double Gaussian given by their equation (7) provides a good fit to HI mapping data.  They have also shown that KISSR 1048 does indeed exhibit lower \hi\ fluxes near the center of the galaxy.  Thus, we follow their beam correction prescription for this particular galaxy and find that $f_c$=1.78, assuming a Gaussian shaped beam with HPBW=3\farcm3.  This yields a corrected flux of 31.61, which is in better agreement with the Haynes et al. (1998) data.  We adopt this value for the corrected flux of KISSR 1048 in the analyses that follow.

\section{Analysis}

\subsection{Derived HI Quantities}

For the analyses that follow, we compute several quantities from the radio data.  These derived quantities are reported in Table 4.  The column entries shown there are as follows:
\renewcommand{\labelenumi}{(\arabic{enumi})}
\begin{enumerate}
\item Galaxy identification as given in the first KISS Red survey strip, repeated from Table 1.
\item HI gas mass of the galaxy, obtained via the standard conversion $M_{HI} = 2.36 \times 10^5 \;D^2 \;F.I._c$, where $D$ is the distance in Mpc, $F.I._c$ is the beam-corrected flux integral, and the resulting gas mass is in solar units.  In the table, M$_{HI}$ is reported in units of $10^8 M_{\odot}$.  Upper limits for the three severely confused sources with neighbors brighter than the targets themselves (KISSR 146, 148, 401) are computed as follows.  In the spectra of KISSR 146 and 148, the measured flux is a combination of emission from three galaxies, KISSR 146, KISSR 148 and UGC 8033, a large spiral galaxy which is most likely producing most of the emission.  As a conservative upper limit, we compute the HI mass for both KISSR 146 and 148 from one-third of the observed flux (given in Footnote 1 of Table 1) from the confused region.  KISSR 401 is confused with two other KISS galaxies which are not in the low luminosity sample, KISS 399 and 400.  Since the target is the weakest optical source in the trio, our listed upper-limit HI mass is computed from one-third of the observed flux (given in Footnote 2 of Table 1) from the confused region.  Upper limits for non-detections and the remaining 7 confused sources are computed using the upper limit fluxes discussed in $\S$ 3.3 and given in Table 1. 
\item Ratio of HI gas mass to $B$-band luminosity, M$_{HI}/L_B$ in units of M$_{\odot}$/L$_{\odot}$.  
\item The present star formation rate (SFR) of the galaxies computed via:
\begin{equation}
SFR \:\: [M_{\sun}/yr] = 7.9 \times 10^{-42} L_{H\alpha}  
\end{equation}
Kennicutt (1998), where $L_{H\alpha}$ is the luminosity in the $H\alpha$ emission-line in units of ergs s$^{-1}$.  Details of the calculation of both this quantity and of $L_{H\alpha}$ are given in $\S$ 4.4.  This relation assumes solar metallicity, a Salpeter IMF and masses between 0.1 M$_{\sun}$ and 100 M$_{\sun}$.  $SFR(z_{\odot})$ is given in units of LOG(M$_{\sun}/$yr.)
\item A metallicity dependent, present star formation rate of the galaxy, $SFR(z)$, based on Starburst 99 models (Leitherer et al. 1999; see $\S$ 4.4 for more details).  $SFR(z)$ is given in units of LOG(M$_{\sun}/$yr.)
\item The gas depletion timescale based on a solar-metallicity SFR, $\tau_{\sun}=M_{HI}/SFR_{\sun}$.  $\tau_{\sun}$ is reported in units of LOG(yrs).
\item The gas depletion timescale based on a metallicity dependent SFR, $\tau_{z}=M_{HI}/SFR_{z}$.  $\tau_{\sun}$ is reported in units of LOG(yrs).
\end{enumerate}

Note that dynamical masses, based on the width of the HI line profile, currently cannot be estimated.  This is because axial ratios cannot be accurately measured for the majority of the galaxies, due to the coarse resolution of the available survey images (2\farcs03 pix$^{-1}$), and the small angular size of the sources. 

\subsection{HI Line Widths}

As illustrated in the spectra shown in Figure 2, the KISS low luminosity galaxies exhibit a wide variety of HI line profile shapes.  Although there are many examples of narrow, single-peaked 21-cm lines typical of low-mass dwarfs, there are also instances of broad, double-horned profiles which are more characteristic of higher mass spiral galaxies.  This variety is reflected in the large range of line widths present in the sample (Figure 5; also see the distribution of HI masses in Figure 13).  The distribution of line widths (measured at 50\% of the peak) extends from 39 km s$^{-1}$ to 311 km s$^{-1}$ and has a median of 132 km s$^{-1}$ and a mean of 139 km s$^{-1}$.  Widths of profiles belonging to targets which have been determined to be confused with neighboring objects (given in parentheses in Table 1) have been excluded from the computation of these statistics.  The distribution in Figure 5 is not what one would expect for a pure dwarf galaxy sample, which should have smaller line widths on average.  For example, the sample of 88 BCGs observed by Thuan et al. (1999) has a mean $w_{50}$ of 92 km s$^{-1}$ with a maximum of 160 km s$^{-1}$, while the composite sample of 36 low luminosity LSBGs and BCGs observed by Staveley-Smith et al. (1992) has a median $w_{50}$ of 84 km s$^{-1}$ with a maximum of 218 km s$^{-1}$.  The 139 dwarf galaxies in the composite sample of Salzer et al. (2002) also have narrow line widths -- the median $\Delta$w$_{50}$ is 88 \kms with 75\% of the galaxies having widths below 120 \kms.  In the present KISS HI sample, Figure 4 illustrates that the larger HI widths tend to belong to galaxies at the upper end of the luminosity distribution (shown in Figure 1).  It would thus be reasonable to suspect that imposing a fainter luminosity limit would lower both the average and maximum widths, especially as no corrections for internal absorption have been applied and more massive, heavily extincted galaxies could be present in our sample.  This is indeed the case -- excluding the 47 galaxies with $M_B < -17.0$ results in a sample with average, median and maximum widths of 112, 99 and 224 \kms respectively. 

However, Figure 4 also shows that there are galaxies which are luminous ($M_B < -17.0$) but possess fairly narrow widths.  Although this can be characteristic of small, face-on spirals, it can also be characteristic of dwarf galaxies that are undergoing a major starburst event.  Thus, simply lowering the absolute magnitude limit of the sample will not yield a comprehensive list of the dwarf galaxies in KISS, but will likely exclude strongly bursting low-mass objects such as the BCDs.  Thus, simply lowering the absolute magnitude limit of the sample will not yield a comprehensive list of the dwarf galaxies in KISS, but will likely exclude strongly bursting low-mass objects such as the BCDs.  Since our goal here is to provide a complete sample of dwarf {\it candidates} for more detailed, follow-up investigations, and not to produce a pure dwarf galaxy sample {\it per se}, in the remainder of the paper we report global HI properties for the entire ($M_B < -18.0$) sample only.  Those seeking to use this sample for further studies of dwarf galaxies may wish to impose additional selection criteria, such as an upper line-width cut-off in conjunction with the morphological appearance of the galaxies in the survey images (Figure 3), to exclude more massive galaxies with large rotational velocities and other non-dwarfs from the sample.

Plotting the data in the log($w_{50}$)-log($M_B$) plane (Figure 4) also yields the Tully-Fisher relation for this sample.  A linear least-squares fit produces the relation $\Delta w_{50} \propto (L_{B}/10^9 M_{\sun})^{\alpha}$, where $\alpha = 0.285 \pm 0.049$ (solid line).  The fairly large scatter can be reduced by correcting the HI linewidths for inclination, although probably not to the level seen in the relation for spiral galaxies.  This is because many of the objects in our sample have irregular morphologies, so that the meaning of the inclination itself is ill-defined.  Even with the scatter present in Figure 4 however, the Spearman rank-order correlation coefficient is 0.45 which is significant at the 99.998\% confidence level for a sample of this size.  When the data points from the 24 possibly confused sources are removed and the fit is reevaluated, the resulting power-law slope agrees with the fit to the entire data set to within 1$\sigma$.  

Finally, it is interesting to note that the observed line widths do not continue down to arbitrarily low values, which is in contrast to studies of quiescent, low-mass dIrr galaxies (e.g. Eder \& Schombert 2000).  This effect is also seen in Salzer et al. (2002), who examined the HI properties of star-forming dwarfs primarily from color and/or line-selected surveys such as the Markarian, Michigan, Case and Wasilewski catalogs, and observe a sharp fall-off in the number of galaxies with widths below 40 \kms.  A similar drop can been seen in our distribution within the same width regime.  We reiterate the suggestions made in Salzer et al. (2002) that this apparent width threshold may represent a physical limit corresponding to (1) the point below which galaxies do not possess densities sufficient to initiate a global star-formation episode strong enough for the object to be classified as a starburst galaxy in line or color-selected surveys, and/or (2) feedback processes such as stellar winds and supernova heating that inject energy into the ISM.  For the KISS low luminosity sample, current star formation rates are detected down to 10$^{-3}$ M$_{\odot}$ yr$^{-1}$, with an abrupt drop in the number observed at 10$^{-2}$ M$_{\odot}$ yr$^{-1}$ (see $\S$ 4.4).  The three objects with widths below 45 \kms\ have SFRs between 0.1 and 1 M$_{\odot}$ yr$^{-1}$.

\subsection{Correlations with $M_{HI}/L_B$}

The distribution of HI mass to blue luminosity ($M_{HI}/L_B$) for this sample is shown in Figure 6.  The range of $M_{HI}/L_B$ is the same as that seen in previous observations of dwarf and other late-type galaxies, such as those by Staveley-Smith et al. (1992).  Our sample includes both gas poor objects with log$(M_{HI}/L_B) < -0.5$ and gas rich objects with  log$(M_{HI}/L_B) > 0 $.  For comparison, the average values and standard deviations of log$(M_{HI}/L_B)$ for different morphological groupings from Haynes and Giovanelli (1984) are over-plotted.  The mean and standard deviation of the full sample (N = 109) are $-$0.08 and 0.35 respectively, which is consistent with Haynes and Giovanelli's result of $-$0.04 and 0.33 for galaxies which are later than Sc.  Since upper-limit values for confused and non-detected sources are included, our reported full-sample mean will be higher than the true mean.  When the upper-limit values are removed, the mean drops to $-$0.13 (with $\sigma$ = 0.33).  

Since gas-richness can be an indicator of the evolutionary state of an object, with more gas-rich galaxies tending to be less evolved, it is interesting to investigate the relationships between $M_{HI}/L_B$ and other galaxy characteristics.  Here we present the $(M_{HI}/L_B)-L_B$ and $(M_{HI}/L_B)-$metallicity correlations for this sample.

First, we report a weak anti-correlation between $M_{HI}/L_B$ and luminosity.  In Figure 7, $M_{HI}/L_B$ is plotted against $M_B$ and a least-squares fit to the full sample is shown.  The solid line shows a linear least squares fit to the detections (solid symbols) and corresponds to $M_{HI}/L_B \propto L_B^{\beta}$ where $\beta = -0.2 \pm 0.1$.  The Spearman rank-order correlation coefficient for these data is 0.22, which is significant at the 95\% confidence level.  Including the upper-limit values in the fit yields a power law slope of $-0.3 \pm 0.1$, which is within the errors of the first result.  Recomputing the fit with the 24 possibly confused galaxies removed also produces a power-law slope consistent to within 1$\sigma$ of the first result.  The weakness of the result is illustrated by re-evaluating the fit and statistics when the two faintest points are removed from the sample.  This exercise produces a slope of $-0.15 \pm 0.11$ and a correlation coefficient of 0.14.  The relationship between these quantities has been previously well-studied.  The results agree within 3$\sigma$: Staveley-Smith et al. (1992) find a slope $-0.3 \pm 0.1$ for a sample of LSB and BCD galaxies,  Smoker et al. (2000) find $\beta = -0.2 \pm 0.1$ for a subsample of the University of Michigan dwarf emission-line galaxies and Davies et al. (2001) find $\beta = -0.4 \pm 0.1$ for the HI Parkes All Sky Survey galaxies).  Thuan \& Seitzer's (1979a) sample of 145 UGC dwarfs and Salzer et al.'s (2002) sample of 139 dwarfs cataloged in various objective-prism surveys for UV-excess or emission-line galaxies find no statistically significant trend.  Our results are statistically consistent with all of these studies.  The variation in the derived values for $\beta$ are likely to be primarily caused by sample differences.

Second, we also find a weak relationship between the metallicities and $M_{HI}/L_B$'s of our galaxies, with the metal abundance falling as the gas richness increases (Figure 8).  The metallicities used here are coarse estimates and have errors of 0.15 dex as discussed in detail in Melbourne \& Salzer (2002).  We note that the three most metal-poor galaxies (12 + log(O/H)$<7.5$) in this sample have an additional {\it systematic} uncertainty associated with their reported metallicities.  This comes about because the abundance estimates for the most metal-poor objects are derived solely from the [NII]/H$\alpha$ ratio, but the weakness of the [NII] line makes the measurement difficult for the most extreme objects.  This additional error is limited to the handful of most metal deficient systems.  The nature of this error is to underestimate the true metal abundance up to 0.3 dex.  A linear least squares fit to the detections (solid symbols) in Figure 8 yields  12 + log(O/H) $\propto (M_{HI}/L_B)^{0.2 \pm 0.1}$.  The Spearman rank-order correlation coefficient for these data is 0.22, which is significant at the 97\% confidence level.  Again however, deleting the two most gas-poor data points weakens the correlation considerably: the slope and correlation coefficient become $-0.17 \pm 0.13$ and 0.15, respectively.  Although it is clear that most of the possibly confused galaxies fall below the best-fit line, we do not recompute the fit with these 24 galaxies removed since they do not appear to be biased in the HI-related variable, $M_{HI}/L_{B}$.  Smoker et al. (2000) also note a correlation between metallicity and gas-richness, but this result is only based on a sample of 15 galaxies so the authors do not attempt a formal fit.  

In considering these relationships however, one must keep in mind that $L_B$ is being used as an observable indicator for the more fundamental quantity of mass, and that the light in the B-band is strongly affected by star-formation.  This is particularly true for low luminosity galaxies where sites of recent star formation may only be a small fraction of the stellar mass, but contribute a much larger fraction of the total light than in high-luminosity objects.  In dwarf starbursting galaxies, the optical luminosity increases by an average of 0.75 magnitudes during a star-formation event (Salzer \& Norton 1999), leading to a 0.3 dex average decrease in $M_{HI}/L_B$.  So although the general characteristics of the class of low luminosity galaxies can be established with $L_B$ and $M_{HI}/L_B$, similar analyses with observable quantities more representative of the mass are needed to investigate the specific relationships between the different morphological classes of dwarf galaxies.  

A clearer picture may be gained by repeating the above analyses with near-infrared or redder photometry, since the light in these wavebands is less affected by recent star-formation and dust, and is a better measure of the total stellar mass.  Two examples of this are in the papers by Schombert, McGaugh \& Eder (2001) who examine the relationship between $M_{HI}/L_B$ and  $M_B$ in the I-band for low surface brightness dwarf galaxies from the Second Palomar Sky Survey (Reid et al. 1991), and Boselli et al. (2001) who study it using H-band luminosities for a sample of late-type galaxies taken from the Zwicky Catalog (Zwicky et al. 1961) and the Virgo Cluster Catalog (Binggeli, Sandage \& Tammann 1985).  Both groups include spirals in their analyses for comparison.  Over a broad baseline ($-24 < M_I < -11$ and $10^8 < \mbox{log}(L_H) < 10^{11.5}$) there is a clear trend of increasing gas richness with decreasing luminosity, but these two variables exhibit only a weak correlation when the lowest luminosity galaxies are considered by themselves.  This is consistent with our conclusion using B-band photometry.  We are currently obtaining I-band and near-IR imaging for this sample of dwarf star-forming galaxies and will re-examine these relationships using these data in a future paper. 

We have also made comparisons with the chemical evolution models of Ferrara \& Tolstoy (2000; hereafter F\&T), who incorporate the dynamical effects of dark matter (assuming a modified isothermal sphere density distribution for the dark halo) into their calculations.  In Figure 9 we show the metallicity of our galaxies plotted against their HI masses (filled circles).  Galaxies that only have upper-limits on their HI masses are excluded.  Overplotted are lines of constant non-baryonic to baryonic (gas + stars) mass ratios ($\phi$) from the F\&T models (see F\&T's Figure 4).  Also plotted are the isolated, quiescent, low surface brightness galaxies (LSBGs) from van Zee et al. (1997a,b) (open stars).  Note that the KISS low luminosity galaxies sample quite a large range of $\phi$, while the van Zee LSBs preferentially cluster around small $\phi$.  This is reasonable given that the KISS galaxies are a much more heterogeneous collection of objects than the van Zee galaxies.   Since the KISS sample is more strongly star-forming, this is also consistent with F\&T's framework where larger dark matter haloes with greater central densities induce larger central SFR densities via the Schmidt law.  If the models are correct, and we assume that $\phi$ is relatively constant with time (galaxies evolve along lines of constant $\phi$), then the class of galaxies that the van Zee LSBGs represent can at most be progenitors of only a small sub-sample of bursting galaxies that have similar dark matter fractional content, i.e. the van Zee LSBGs cannot simply transform themselves into any type of starbursting dwarf.  Similarly, the faded counterparts of the KISS galaxies should be dim objects with higher $\phi$.  The structural parameters of LSBGs and KISS galaxies with similar $\phi$ can be compared to further test these assertions.  Locating the sub-population of Blue Compact Dwarfs within this diagram would also be interesting, since they would be expected to have large dark matter fractions.  Observational evidence that the {\it baryons} in BCDs are more centrally concentrated than in LSB dI galaxies has already surfaced.  Papaderos et al. (1996) and Salzer \& Norton (1998) have shown that when compared with a dI or dE at equal B-band luminosities, the underlying component of a BCD has a central surface brightness brighter than about 1.5 magnitudes and an exponential scale length smaller by a factor of $\sim$2.  Understanding the differences in the distribution of the dark component for the different morphological types of dwarfs will require high-resolution HI mapping as well as higher resolution optical and near-IR images to assess the morphologies of the galaxies in the sample.

\subsection{Star Formation Rates and Gas Depletion Time Scales}

The star formation rates (SFRs) given in Table 4 are calculated using H$\alpha$ line fluxes measured from the KISS objective-prism spectra.   These fluxes are a better measure of the total H$\alpha$ emission than those derived from slit-spectroscopy, which only samples a fraction of the target galaxy.  Further, spectral line measurements from the objective-prism images are all on precisely the same flux scale.  The calibration of these fluxes is described in detail in KR1. 

Two corrections have been applied to the fluxes, one for the presence of blended [NII] emission, and the other for absorption.  The [NII] correction is:

\begin{equation}
f(\mbox{H}\alpha)_{corr1} = \frac{f(\mbox{H}\alpha)}{1.33(j_{[NII]\lambda6583}/j_{H\alpha}) + 1} ,
\end{equation}

\noindent and the absorption correction is:

\begin{equation}
f(\mbox{H}\alpha)_{corr2} = f(\mbox{H}\alpha)_{corr1} \;10^{0.74 C_{H\beta}} ,
\end{equation}

\noindent where $C_{H\beta}$ is the reddening coefficient derived from Balmer line ratios measured from the follow-up slit spectra.

\noindent Using $f(\mbox{H}\alpha)_{corr2}$, H$\alpha$ luminosities are computed (assuming $H_o$= 75 km s$^{-1}$ Mpc$^{-1}$ and $q_o$ = 0.5), and then converted to current SFRs using the Kennicutt (1998) prescription,

\begin{equation}
SFR(Z_{\odot}) = 7.9 \times 10^{-42} \;L_{H\alpha} \;\;\;\;[M_{\odot} yr^{-1}]
\end{equation}

\noindent which assumes solar metallicity and a Salpeter IMF with masses between 100 and 0.1 M$_{\odot}$.  SFRs computed in this way are listed in column 4 of Table 4.  This linear conversion is a standard SFR estimator that is used when recourse to more detailed stellar synthesis modeling for individual galaxies is not taken or possible.  However, the SFR is dependent on variables such as the assumed initial mass function, the time since the initiation of the star formation event, the star formation history, and the metallicity of the galaxy.  In the following simple exercise, we specifically focus on the impact of the metallicity on the SFR and associated gas depletion timescales for our sample.  We do this to illustrate the general direction and magnitude of the errors that are incurred when the above estimator is applied to low-metallicity galaxies.  

Metallicity will influence the H$\alpha$-based SFR of a galaxy through its effect on stellar opacity and consequently, on the number of ionizing photons produced.  A metal-rich stellar population will contain stars which are both cooler and less luminous than one which is metal-poor, if all other variables, such as the IMF are held constant.  This is because an increased metal abundance leads to a greater opacity which will increase the pressure in a star, so that for a given luminosity, the radius is forced to increase and the effective temperature is forced to decrease.  In particular, a lower temperature in O and early-type B stars will shift the spectral distribution such that fewer ionizing photons are produced.  Thus, metal-poor stars will emit a substantially larger fraction of ionizing photons than higher metallicity stars of the same bolometric luminosity.  Therefore, fewer O and B stars, and consequently, a correspondingly lower total mass of formed stars, are required to produce the same amount of H${\alpha}$ emission in a low metallicity system.  Since low luminosity star-forming galaxies tend to be metal poor, we would expect the above H$\alpha$-based SFR to over-estimate the true SFR for galaxies in this sample. 

To compute an approximate metallicity-based correction to the SFR computed in Eq. 5, we use the Starburst99 models (Leitherer et al. 1999) to determine the ratio of $N_{LyC}(Z)$, the number of Lyman continuum photons produced in an instantaneous burst model with metallicity $Z$, to $N_{LyC}(Z_{\odot})$, the number produced in a model with solar metallicity.  A Salpeter IMF and an upper mass limit of 100 M$_{\odot}$ is assumed.   The product 

\begin{equation}
SFR(Z) = \frac{N_{LyC}(Z_{\odot})}{N_{LyC}(Z)}\;SFR(Z_{\odot}) 
\end{equation}

\noindent is then taken as the metallicity corrected SFR.
 
In Figure 10, $N_{LyC}(Z)/N_{LyC}(Z_{\odot})$ is plotted against 12 + log(O/H) for burst ages of 2.5, 5, 7.5, 10, 12.5 and 15 Myrs.  This range represents the ages during which systems are luminous enough in H$\alpha$ to be observable by KISS, assuming that 0.45 H$\alpha$ photons are produced for every Lyman continuum photon and that the KISS detection limit is nearly zero at $L_{H\alpha} \sim 10^{39} $ergs s$^{-1}$ (Gronwall et al. 2002).  The range of ratios of ionizing photons is largest for the lowest metallicity model, which varies from 1.5 to 9, and smallest for the super-solar model, which varies from 0.4 to 0.6.  We choose to evaluate SFR(Z) at 15 Myr and take the errors in $SFR(Z)$ as the spread in $N_{LyC}(Z_{\odot})/N_{LyC}(Z)$ with age as shown in Figure 10.  Values for $N_{LyC}(Z)/N_{LyC}(Z_{\odot})$ are found by linearly interpolating between the points given by the Starburst99.  Note that we have linearly extrapolated beyond the lowest metallicity model point of $Z$ = 0.001 (12 + log(O/H)= 7.62), to extract correction factors for 3 objects in our sample with 0.0006 $< Z <$  0.001 (i.e. $7.40 < 12+log(O/H) < 7.62$).  This extrapolation appears to be warranted given the overall linear relationship between the variables.  

Both the corrected and uncorrected SFRs are plotted against $M_B$ in Figure 11.  As expected, the metallicity correction is such that $ SFR(Z) < SFR(Z_{\odot})$.  The dotted line shows a linear least-squares fit to the solar metallicity based SFRs, while the solid line shows a fit to the SFRs to which correction have been applied.  When metallicity effects are taken into account, the best-fit line for the sample is depressed by $\sim$0.5 dex.  The strong correlation between the two plotted variables is also expected, since in dwarf starbursting galaxies, newly-formed stars will dominate the integrated blue light, so that $M_B$ is essentially an indicator of the SFR.  This relation will not necessarily hold for more luminous normal galaxies with lower star formation rate densities.  

Finally, the effect of the metallicity on the HI gas depletion timescale is considered.  This quantity indicates how much longer a galaxy can continue to sustain its present SFR before its supply of fuel is exhausted.   The distributions of $\tau$ are shown in Figure 12.  The median of the $\tau(Z_{\odot})$ distribution is 5 Gyr, while that of $\tau(Z)$ is 13 Gyr.  The metallicity given by 12+log(O/H) is over-plotted against $\tau(Z)$ with error bars indicating the possible variation with burst age.  The points and their associated ranges show that (1) the distribution of $\tau(Z)$, when binned by 0.25 dex, does not change significantly when corrections to the SFR are made for burst age, and that (2) small shifts of the peak of the distribution to larger timescales are more probable than shifts to smaller timescales.  Thus, the tendency for metallicity to increase the gas depletion timescale cannot be erased by the variation of $\tau$ with burst age, and the difference between the two distributions seems to be real.

When the metallicity is accounted for in the determination of the star formation rate, a much smaller fraction of this sample has gas depletion timescales which constrain their star-formation histories to be composed of short bursts.  Whereas $\tau(Z_{\odot})$ is smaller than 5 Gyr for half of the sample, the corresponding fraction for $\tau(Z)$ is only about 10\%.  Alternatively, many more galaxies can sustain their present SFRs for another Hubble time.  Although $\tau(Z_{\odot})$ is larger than 10 Gyr for only $\sim$20\% of the sample, it rises to over 70\% when metallicity corrections are applied.  These differences emphasize the importance of accurately determined SFR in the application of gas depletion timescales to our understanding of the evolutionary histories of strongly star-forming dwarf galaxies.  Blue compact dwarfs, as defined by Thuan (1991) as the set of galaxies with $M_B>-18.0$ and HII region-like spectra, have been commonly argued to have star-formation histories comprised of short bursts which last $\sim10^8$ yrs interspersed by quiescent periods of $\sim10^9$ yrs.  This picture was based in large part on the need to make the derived solar-metallicity based SFRs ($\sim0.1 M_{\odot} yr^{-1}$) compatible with the observed HI masses ($\sim10^8 M_{\odot}$).  Our analysis shows that this constraint on the star formation histories is not needed for the overwhelming majority of star-forming dwarf galaxies of the type cataloged by KISS.  In other words, star-formation histories which are less bursty, and involve more nearly continuous star formation can and should be considered for these objects.  Adopting less bursty star-formation histories also reduces the long standing problem of locating the post-burst counterparts of BCDs, which should be very common if the burst duty cycle is as low as 10\%, but are not found in existing galaxy samples.

As argued by van Zee (2001), these longer timescales can be evidence for ``quasi-continuous'' star formation.  However, the quantities as defined above assume that star formation will cease only when {\it all} the HI gas in the galaxy has been processed, and also neglect to account for consumption of molecular gas.  The amount of HI available for star-formation will likely be less than the total HI in the galaxy, since the distribution of HI in low luminosity star-forming galaxies often extends beyond the optical radius, where it exists in a low density state.  Corrections for the amount of HI available for star-formation and the molecular gas will therefore work in opposite directions.  Although a rough estimate of the fraction of HI involved in star-formation can be made from HI synthesis maps (e.g. van Zee et al 1998, 2001), a comparable estimate cannot be made for the molecular hydrogen content.  This is due to the current difficulty of detecting CO in dwarfs and because of our poor knowledge of the CO to H$_2$ conversion factor in these metal deficient systems (e.g. Taylor et al. 1998).  Thus, the degree to which these two factors will offset each other will not be known until our understanding of the H$_2$ content of dwarf galaxies is improved.

\subsection{\hi\ Mass Function for Low Luminosity Star-forming Galaxies}

Since this data set has been cataloged in a uniform way and has well-understood completeness limits, we are able to investigate how the sub-population of low luminosity star-forming galaxies contributes to the overall HI mass function (HIMF).  The HIMF describes the number density of objects with differing neutral hydrogen gas masses as a function of that mass.  Integrating over the HIMF yields the HI density in the local universe ($\Omega_{HI}$), which is an important observational constraint on models of galaxy evolution and cosmology.  The fraction of the HIMF due to low luminosity star-forming galaxies is particularly interesting in light of some reports of a trend of increasing M$_{HI}$/L$_B$ with decreasing L$_B$ (see Section 4.3) coupled with indications of a possible upturn at the low-mass end of the HIMF (Schneider et al. 1998, Rosenberg \& Schneider 2002).

The standard $\Sigma$ (1/$V_{max}$) method (Schmidt 1968) is used to determine the HIMF for the KISS HI sample.  Within each HI mass bin, this technique sums the reciprocals of the volumes corresponding to the maximum distances at which galaxies can be placed and still remain in the sample.  Since we are interested in the HIMF for a specialized subset of the general population, the calculation of V$_{max}$ will be dependent upon the selection function of the survey method which originally produced the subset.  In this case,  emission-line strength is the primary characteristic that determines whether an object is included in KISS, while a velocity limit of 11,000 km s $^{-1}$ determines whether a KISS ELG is included in the HI sample.  Therefore, $V_{max}$ is taken to be the smaller of: (1) the object's limiting volume in KISS, based on $F_{L+C}$, the sum of the H$\alpha$ flux and the flux in the continuum under the line, or (2) the volume corresponding to 11,000 km s $^{-1}$ cut-off. 

Completeness of the sample has been assessed through an $V/V_{max}$ (Schmidt 1968; Huchra \& Sargent 1973) analysis of $F_{L+C}$, which uses the fact that a uniformly distributed sample of objects will have $<V/V_{max}>$=0.5 for all flux levels at which the sample is complete.  This analysis for the KISS galaxies, as well as the computation of limiting volumes in KISS, is fully described in Gronwall et al. (2002).  We briefly summarize the results here.

In Gronwall et al. (2002), KISS is shown to be 100\% complete to $m_{L+C}=14.4$ (for convenience, $F_{L+C}$ is placed on a magnitude scale using the relation $m_{L+C}=-2.5$ log $F_{L+C}+15.0$, where the zero point of the magnitude scale is arbitrary).  $V/V_{max}$ is stable to better than 10\% between $12 < m_{L+C} < 14$, where it varies between 0.6 and 0.55, and it remains close to a value of 0.50 to $m_{L+C}=15.1$, where the completeness of the sample is 70\%.  At magnitudes fainter than 15.1, $V/V_{max}$ and the completeness decrease rapidly.  Galaxies with $m_{L+C}>15.1$ (N=12) have been excluded from the computation of the HIMF.  Corrections for incompleteness have been applied for galaxies with $14.4 \le m_{L+C}<15.1$.  To summarize:
\begin{equation}
V_{max} = \left\{ \begin{array}
{l@{\quad\quad}l}
f_c\:V_{max}(m_{L+C}), & \mbox{if}\quad V_{max}(m_{L+C}) \le V(cz=11,000 \;\mbox{km s}^{-1})\\
V(cz=11,000 \;\mbox{km s} ^{-1}), & \mbox{if}\quad V_{max}(m_{L+C}) > V(cz=11,000 \;\mbox{km s}^{-1})\\
\end{array} \right.
\end{equation}
where $f_c$ is the completeness fraction given in Gronwall et al. 2002, and volumes have been computed based on a survey area of 62.16 square degrees and $q_o=0.5$. 

As a conservative measure, three galaxies with $cz<1000$ \kms (KISSR 73, 314, 1048) have also been excluded from the sample.  This is because radial velocity is not a reliable indicator of distance for such nearby objects since peculiar velocities can be comparable to recessional velocities in the local neighborhood.  These errors in distance can translate into errors in the derived HI mass which are larger than our chosen bin size of $\mbox{log}(\Delta M_{HI}/M_{\odot})= 0.357$ (discussed further below). 

Figure 13 presents our computed HIMFs for the low luminosity KISS HI sample.  The filled squares represent the HIMF calculated from the $\Sigma$ (1/$V_{max}$) method for this sample, where all upper limit HI masses have been included by using the value of the upper limit to determine bin placement.  The error bars show 1-$\sigma$ Poissonian confidence limits.  The bins are chosen in a way that maximizes the number of galaxies within them while still producing a minimum of five evenly spaced HIMF points.  Of course, each of the upper-limits may actually belong to any of the bins below the one in which it was placed.  As a result, the true low-mass end slope may be steeper than implied by the filled squares.  Thus, we have also investigated the extreme case where all the upper-limit data points represent galaxies with HI masses which place them in the lowest mass bin present in the KISS sample.  This second HIMF is represented by the open stars.  Two other HIMFs computed from HI blind surveys are also plotted for comparison.  Open circles show results from Zwaan et al. (1997) while open squares show results from Rosenberg \& Schneider (2002).  The two lower panels show the distribution of HI masses used in the computation of the KISS HIMFs.  The top histogram corresponds to the number of galaxies used to compute each of the points given by the filled squares, and the bottom histogram corresponds to the open stars.  The unshaded portions of these histograms represent upper-limit detections, while the shaded areas represent true detections. 

The fitted curves in the figure are Schechter (1976) functions:
\begin{equation}
\Phi(M) = \frac{dN}{d\;\mbox{log}M} = \Phi_{\ast}\;\,\mbox{ln 10}\;\,\Big(\frac{M_{HI}}{M_{\ast}}\Big)^{\alpha+1}\;\,\mbox{exp}\Big(\frac{-M_{HI}}{M_{\ast}}\Big)
\end{equation} 
where the free parameters are $\alpha$, the slope of the low-mass end, $M_{\ast}$, the characteristic mass that defines the ``knee'' in the curve, and $\Phi_{\ast}$, the normalization factor.  The parameterizations given by the best-fit curves reported in Rosenberg \& Schneider (2002) and Zwaan et al. (1997) are log$(M_{\ast}/M_{\odot}$)=9.88, $\Phi_{\ast}$=0.005 Mpc$^{-3}$, and $\alpha = -$1.53 and log$(M_{\ast}/M_{\odot}$)=9.80, $\Phi_{\ast}$=0.0059 Mpc$^{-3}$, and $\alpha = -$1.2, respectively.  

Minimizing $\chi^2$ for the filled squares yields log$(M_{\ast}/M_{\odot}$)=8.96, $\Phi_{\ast}$=0.0087, and $\alpha$=-0.60 (solid curve).  This HIMF has a rapidly decreasing low-mass slope, especially when compared with the values of $\alpha$ derived from the two HI blind surveys cited above.  Again however, the true slope may be considerably steeper than -0.60 since galaxies with upper-limit detections may actually belong to any one of the lower mass bins.  Repeating the fitting procedure for the stars (where all of the upper-limits have been deposited in the lowest mass bin), gives a Schechter function with a larger slope, as expected (log$(M_{\ast}/M_{\odot}$)=9.09, $\Phi_{\ast}$=0.0046 Mpc$^{-3}$, and $\alpha = -$0.94; dotted curve).  However, this HIMF is still is not as steep as the HIMF predicted for the overall population of galaxies.  Finally, noting that fit to the stars is pulled down by the point at log(M$_{HI}$/M$_{\odot}$)=8.42, we have also performed the fit omitting this particular point so that the steepest possible $\alpha$ permitted by the data may be found.  Even in this most extreme case $\alpha$ does not rise above $-$1.06 (with log$(M_{\ast}/M_{\odot}$)=9.05 and $\Phi_{\ast}$=0.0063 Mpc$^{-3}$; dot-dashed curve).  Thus, from this discussion it is clear that if the HIMF of the universe does indeed have a steeply rising low-mass slope, it is not caused by a large population of low luminosity star-forming galaxies.

Still, we may consider the potential effect of the 12 galaxies with $m_{L+C}>15.1$ that have been excluded from the analysis.  These galaxies are KISSR 55, 61, 85, 97, 105, 120, 193, 471, 856, 1014, 1091, and 1112.  For this set, the average and median $M_B$ (-15.81 and -15.97 respectively) are lower by about a magnitude than the parent sample, which is not surprising.  Since the HI mass distribution of the 12 galaxies is similar to that of the complete sub-sample ($\langle\mbox{log}M_{HI}/M_{\odot}\rangle$ of 8.6 versus 8.8), the lower blue luminosities lead to a slightly higher average gas-richness ($\langle\mbox{log}M_{HI}/M_{\odot}\rangle=-0.007$ compared to -0.13).  These trends are consistent with the weak correlation between $M_{HI}/L_{B}$ and $L_B$ shown in $\S$4.3.  From this cursory analysis, one would expect that the shape of the HIMF given above would not appreciably change even if it were computed using data from a deeper emission-line survey.  We note, however, that when the HIMF is recomputed by including the 10 galaxies with $m_{L+C}>15.1$ which have HI detections, the low-mass slope does become significantly steeper with $\alpha\sim-1.5$.  Because of the small volumes associated with many of the faint $m_{L+C}$ galaxies and the large incompleteness correction factors that are required, this result is inconclusive.  As always, newer, deeper, statistically complete surveys will test the worked based on its shallower predecessors.

To find the contribution of the KISS HI sample to the total HIMF, we first integrate over all masses:
\begin{equation}
\int_0^{\infty} M\Phi(M)dM = \Phi_{\ast}\Gamma(\alpha+2)M_{\ast}
\end{equation} 

\noindent Using the parameters given by solid curve in Figure 13, we find $\rho_{HI}$(KISS M$_B > 18.0)=7.0 \times 10^6 \; M_{\odot}$ Mpc$^{-3}$, or $\Omega_{HI} = 4.5 \times 10^{-5}$, with a statistical error of 20\%.  Results for the parameters given by the dashed and the dot-dashed curves are $\rho_{HI}=5.5 \times 10^6 \; M_{\odot}$ Mpc$^{-3}$ ($\Omega_{HI} = 3.5 \times 10^{-5}$) and $\rho_{HI}=7.4 \times 10^6 \; M_{\odot}$ Mpc$^{-3}$ ($\Omega_{HI} = 4.7 \times 10^{-5}$) respectively.

Comparison with the total HI gas density computed from the HI blind survey from Zwaan et al. (1997), shows that the low luminosity star-forming galaxies found in KISS contain about 15\% of the overall neutral hydrogen in the universe.  This percentage decreases to 10\% if the total HI gas density is instead computed from the Rosenberg \& Schneider HIMF.  

Direct comparison of the number densities given by the HIMFs over a range confined to lower masses yields one measure of the fraction of dwarf galaxies currently undergoing an episode of strong star formation.  First, let us restrict our attention to $10^8<M_{HI}/M_{\odot}<10^9$.  Comparison with the number densities from the Zwaan et al. (1997) HIMF implies that nearly one out of every two dwarf galaxies with HI masses in this range is bursting.  If number densities from the Rosenberg \& Schneider (2002) HIMF are instead used, the ratio decreases to about 1 out of 4.  Extrapolation of the Schechter functions to lower masses implies that the fraction of bursting dwarfs decreases with decreasing HI mass.  This is consistent with the result that gas richness ($M_{HI}/L_B$) does not increase at a significant level with decreasing galaxy luminosity.  There is a large variation in the {\it rates} of decline implied by the different HIMFs.  If the solid KISS HIMF is taken relative to the Rosenberg \& Schneider (2002) HIMF, the fraction of the population which is strongly star forming is already only 1\% at $10^7 M_{\odot}$.  But if the dot-dashed KISS HIMF is taken relative to the Zwaan et al. (1997) HIMF, the fraction is still $\sim50\%$ at $10^7 M_{\odot}$.  The burst fraction at masses below $10^8 M_{\odot}$ can only be clarified when statistics at the low-mass end of the overall HIMF are improved.  As noted by Rosenberg \& Schneider (2002), this will not be resolved within the context of blind HI surveys in the near future because of the practical observational limits of currently available facilities.  However, we will be able to gradually improve statistics for the specialized population of low HI-mass star-forming galaxies.  Since star-forming systems with $M_{HI}/M_{\odot}<10^8$ are intrinsically rare and can only be detected if they are nearby, large areas of sky must be observed to find these galaxies.  It is thus reasonable that we did not find any extremely low HI mass objects within the small swath covered by KR1 (62.2 deg$^{2}$).  The completion of additional KISS fields and deep HI observations of the nearest and faintest emission-line galaxies found in these areas will help to put more stringent constraints on the incidence of star-forming systems at low HI masses.  
\section{Summary \& Conclusions}

A complete sample of 109 low luminosity ($M_B>-18.0$), nearby ($cz<11,000$ km s$^{-1}$), H$\alpha$-selected star-forming galaxies from the KISS catalog have been observed at 21-cm.  Our detection rate is 89\% (97/109).  By examining the KISS composite $B$ and $V$-band survey images, we have found that 9\% (10/109) have companions of comparable or greater optical brightness within the $\sim3\farcm5$ Arecibo beam.  We find that our non-detections and upper-limits for confused sources do not bias the sample in terms of $L_B$ or $M_{HI}$.  The HI properties of this sample are as follows:

\begin{enumerate} 
\item Our sample includes true dwarf galaxies as well as larger but heavily-extincted edge-on spiral galaxies.  This is reflected in our broad distribution of HI line width (39 km s$^{-1}$ to 311 km s$^{-1}$, median=132 km s$^{-1}$, mean=139 km s$^{-1}$, uncorrected for inclination).
\item The range of HI gas richness for this sample (as defined by $M_{HI}/L_B$) is the same as in previous HI surveys of late-type galaxies.  We report weak anti-correlations between the gas-richness and metallicity, and the gas-richness and blue luminosity.  This is consistent with previous results for different samples of dwarf galaxies. 
\item Using the models of Ferrara \& Tolstoy (2000), our galaxies are shown to have a large range of dark-to-visible mass fractions (0 $<M_{dark}/M_{stars+gas}< $300).
\item The median HI gas depletion timescale for this sample increases from $\sim$5 Gyr to $\sim$13 Gyr when a metallicity-dependent SFR is used to calculate timescales instead of a solar-metallicity SFR.  Accounting for metallicity is important since the galaxies in this sample have low luminosities, and tend to be metal-poor.  One interpretation of these statistics is that these galaxies will not deplete their gas supplies for another Hubble time, if they continue to form stars at their current rate.  These results also lift the requirement that BCDs have star formation histories dominated by short bursts of activity, and show that a more nearly continuous mode of star formation is possible. 

\item By computing an HI mass function for this sample, it is shown that the low luminosity star-forming galaxies in KISS contain 10-15\% of the overall neutral hydrogen in the universe.  We find that $\rho_{HI}(\mbox{KISS}\; M_B >-18.0)=7.0\times 10^6 \; M_{\odot}$ Mpc$^{-3}$, or $\Omega_{HI} = 4.5 \times 10^{-5}$ with a $\sim20\%$ statistical error.  The HIMF of this sub-population does not exhibit a steeply rising slope at low-masses.  This is consistent with the result that gas richness ($M_{HI}/L_B$) does not increase at a significant level with decreasing galaxy luminosity.  In the range $10^8<M_{HI}/M_{\odot}<10^9$, we find that 25-50\% of all galaxies are currently undergoing a strong episode of star formation.

\end{enumerate}

\acknowledgments

We would like to thank the Arecibo scientific and technical staff, particularly Tapasi Ghosh, Karen O'Neil and Chris Salter, for their observational and data reduction support.  JCL also acknowledges fruitful interactions with Rob Kennicutt, Jessica Rosenberg, Andrea Ferrara, and Claus Leitherer.  Financial support for this project was provided by NAIC, an NSF Presidential Faculty Award to JJS (NSF AST 95-53020), and a graduate fellowship to JCL through the UA/NASA Space Grant Program.  This work has made use of the NASA/IPAC extragalactic database (NED), which is operated by the Jet Propulsion Laboratory, Caltech, under contract with NASA.




\clearpage

\begin{figure}
\plotone{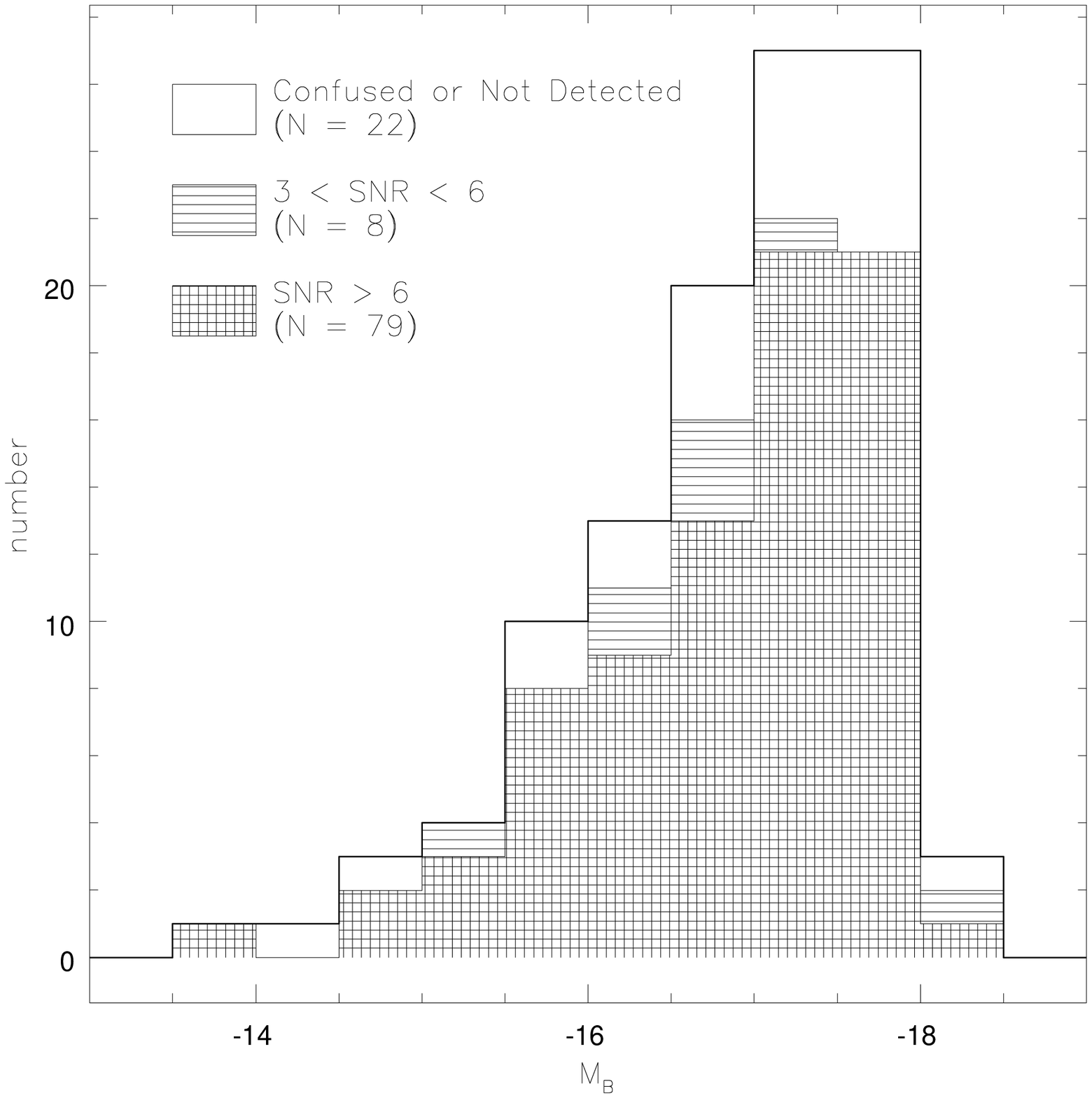}
\figcaption{Distribution of the blue absolute magnitude for this sample.  Note that weak and undetected lines are evenly distributed, so that the results of the analyses will not be biased with respect to $M_B$. }
\end{figure}

\clearpage
\begin{figure}
\plotone{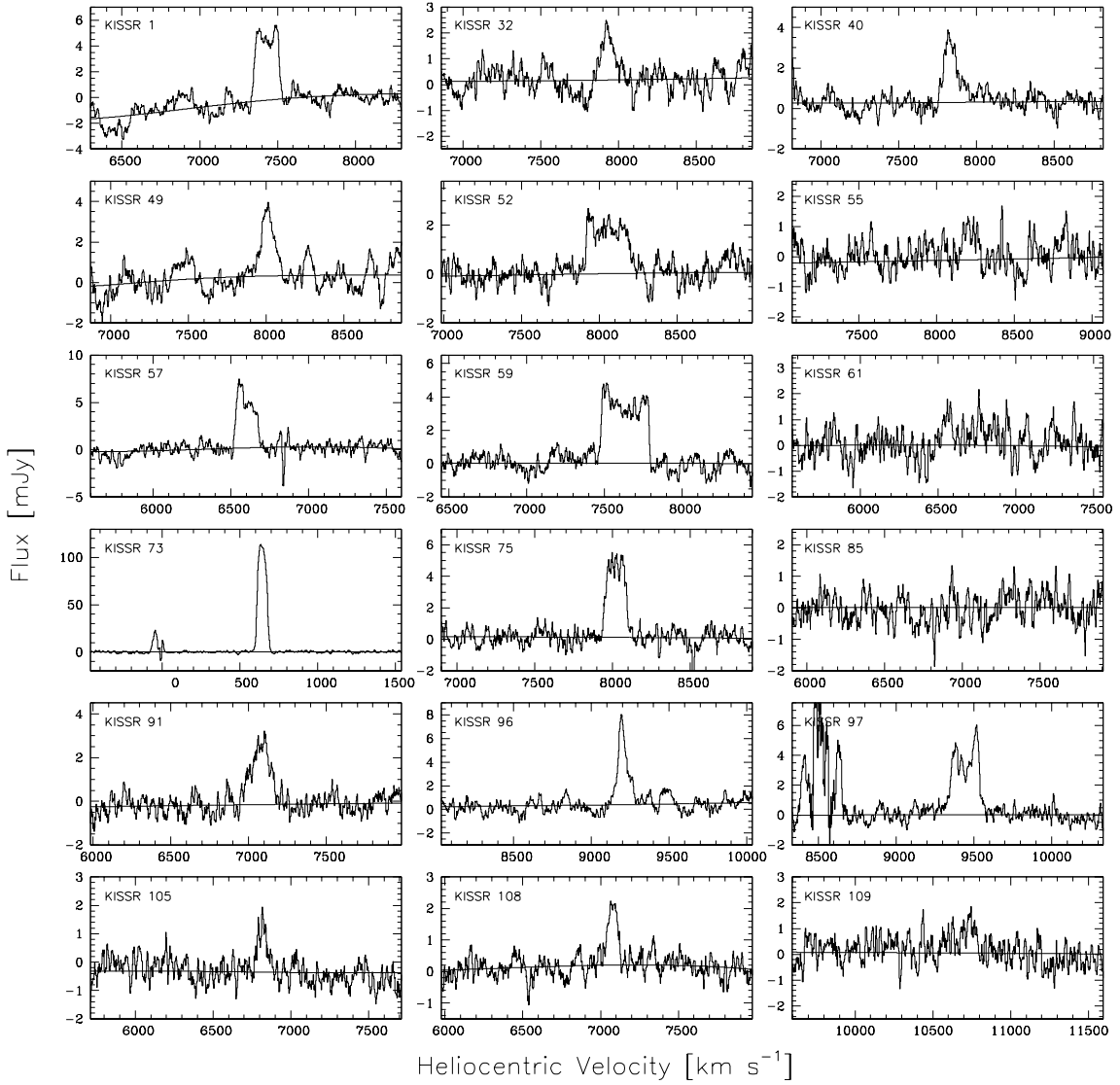}
\figcaption{25 Mhz bandwidth HI spectra for the KISS low luminosity sample.  The channel spacing is 2.6 km s$^{-1}$.  Five-channel boxcar smoothing is applied and polynomial fits to the baseline are shown.  Only the first page is shown.  Contact jlee@as.arizona.edu to obtain the complete set of spectral data plots.}
\end{figure}

\begin{figure}
\plotone{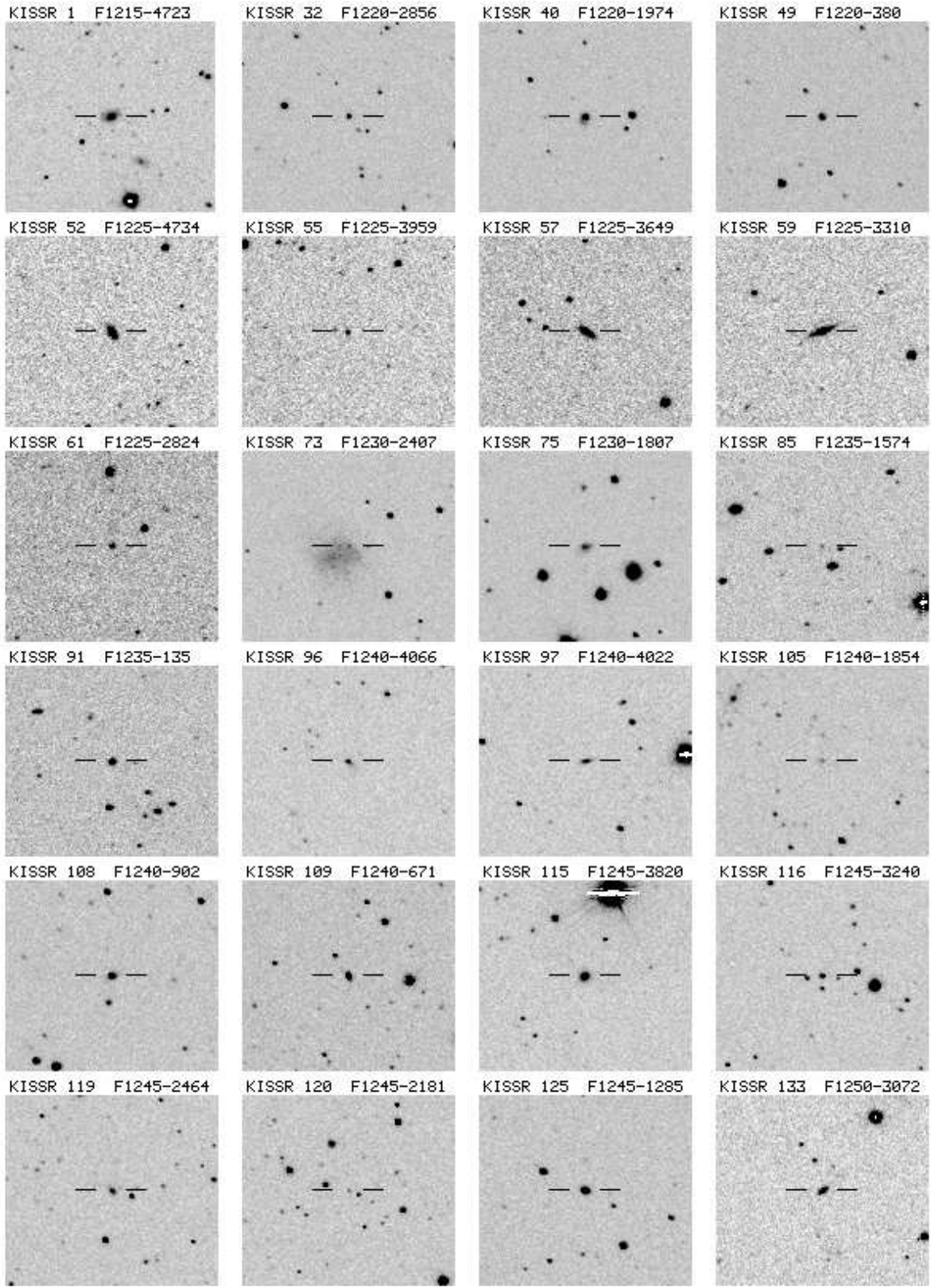}
\figcaption{$4\farcm5 \times 4\farcm0$ combined ($B+V$) direct images of galaxies in the KISS low luminosity sample.  North is towards the top of the page and east is towards the left.  Only the first page is shown.  Contact jlee@as.arizona.edu to obtain the complete set of optical images.}
\end{figure}

\begin{figure}
\plotone{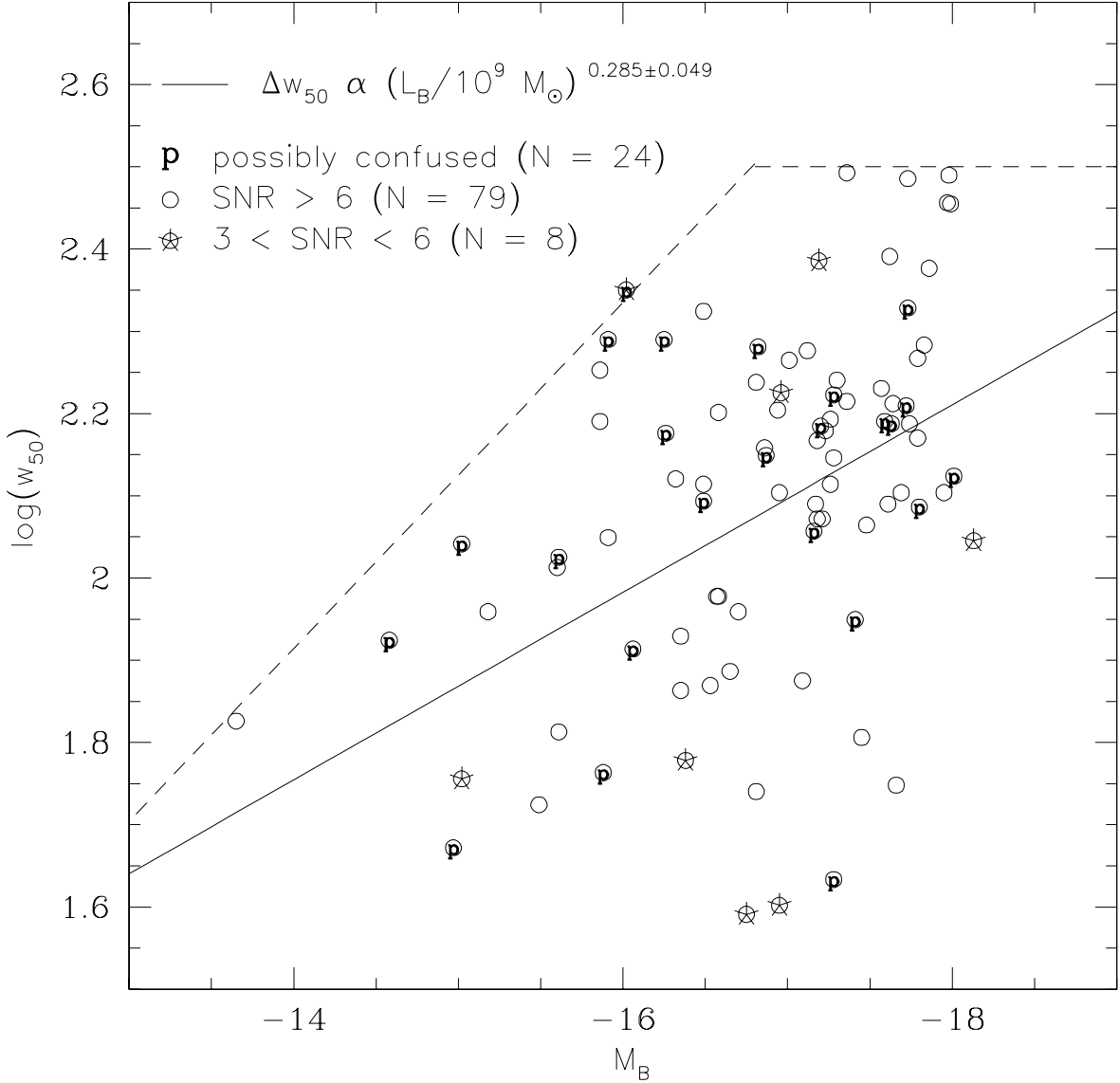}
\figcaption{Velocity width measured at 50\% of the peak flux plotted against blue luminosity.  The letter ``p'' is used to denote symbols which represent galaxies that are possibly, but unlikely to be, confused (see $\S3.2$).  Severely confused data points have been omitted.  The dashed line outlines the upper envelope of observed widths in this sample.  Upper-limit fluxes are calculated using widths corresponding to the objects' luminosities given by these dashed lines.  The solid line shows a linear least-squares fit to all of the plotted data, which shows a trend of increasing line-width with increasing luminosity.  This plot essentially shows the Tully-Fisher relation for the KISS low luminosity star-forming sample.}
\end{figure}

\begin{figure}
\plotone{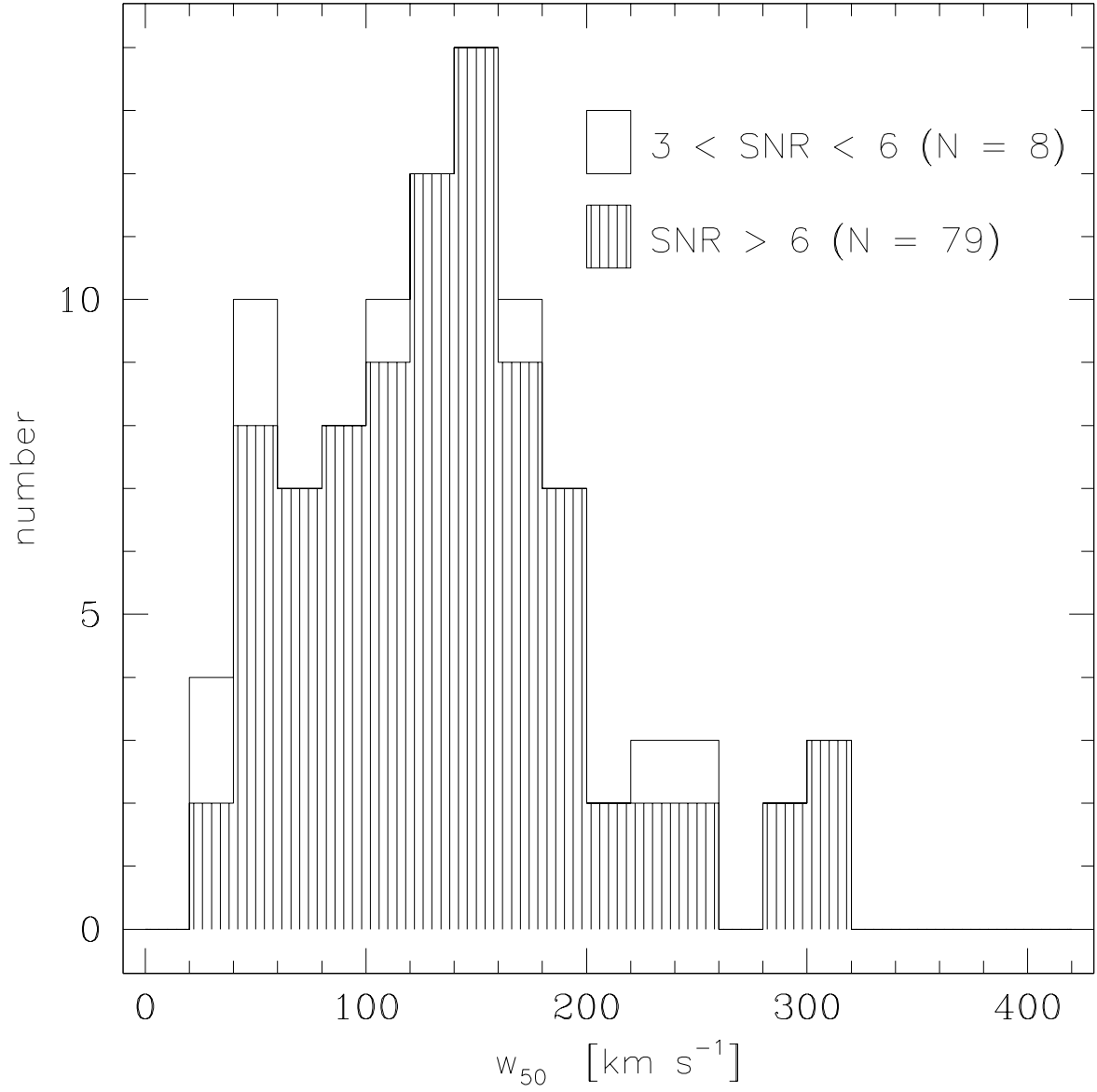}
\figcaption{Distribution of velocity width measured at 50\% of the peak flux.  The median and average widths are 132 km s$^{-1}$ and 139 km s$^{-1}$, respectively.}
\end{figure}

\begin{figure}
\plotone{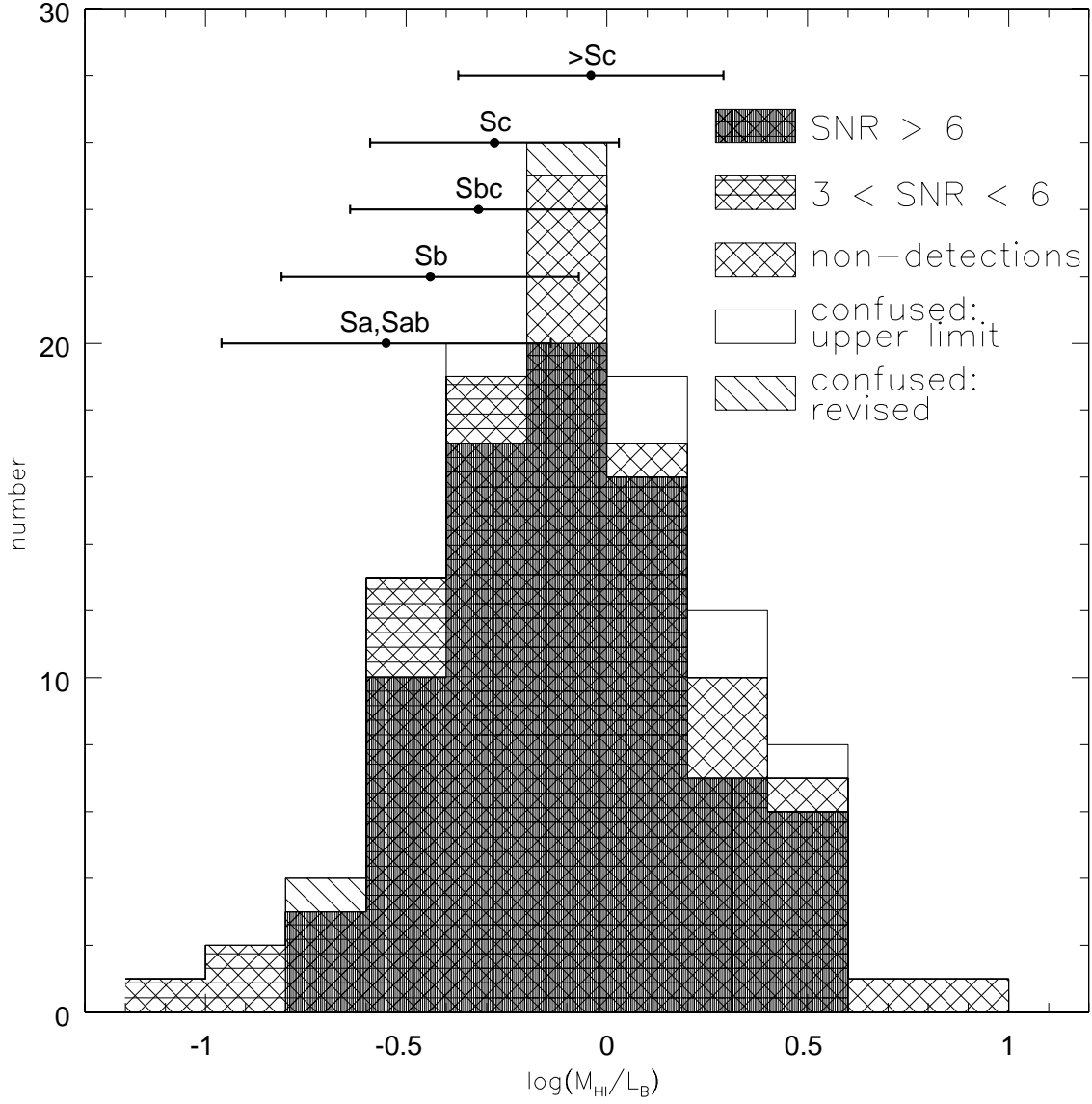}
\figcaption{Distribution of the HI mass to blue luminosity. Average values and standard deviations of log$(M_{HI}/L_B)$ for different morphological groupings from Haynes and Giovanelli (1984) are over-plotted.}
\end{figure}

\begin{figure}
\plotone{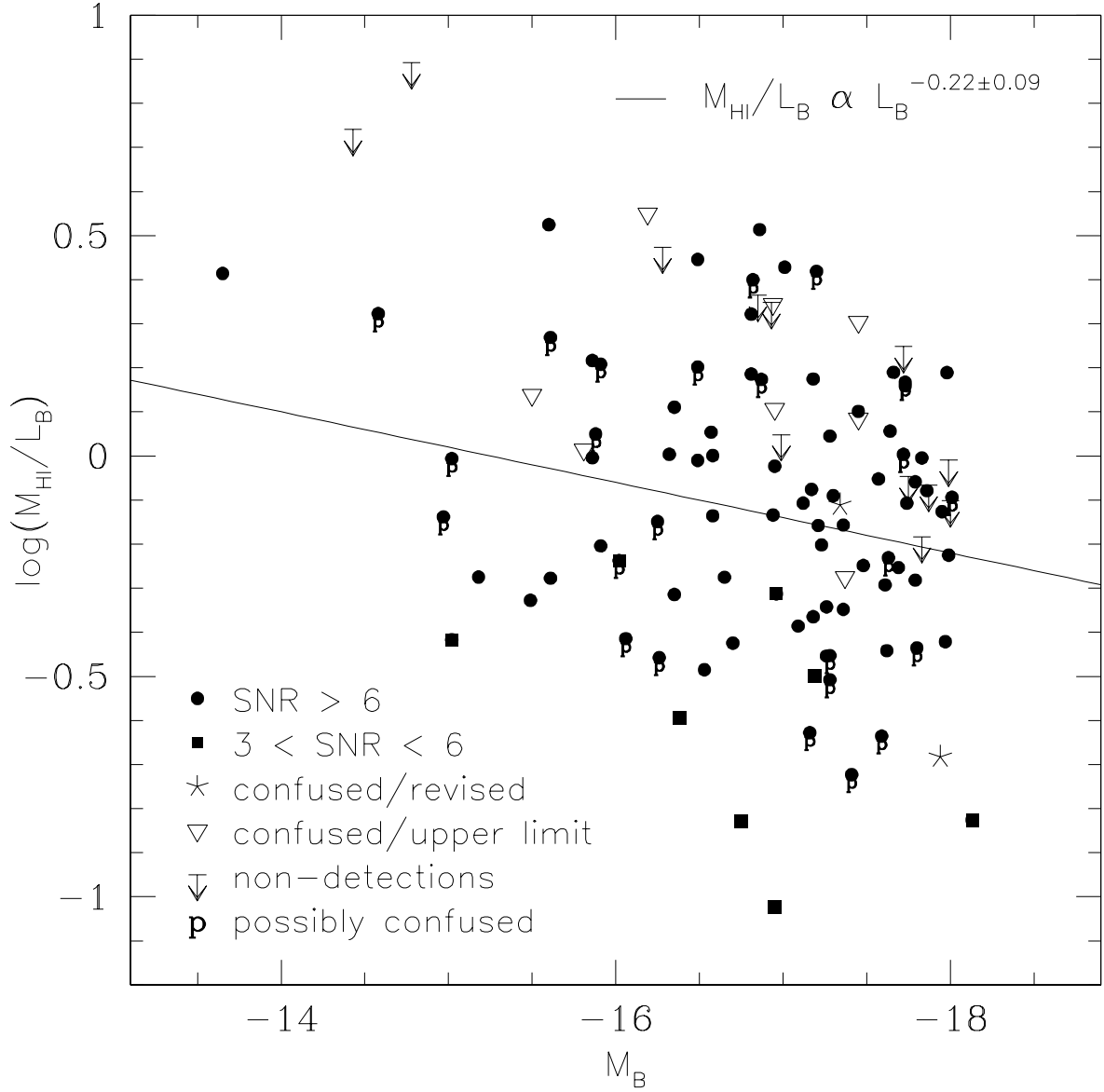}
\figcaption{Gas richness measured by the ratio of HI mass to blue luminosity plotted against blue absolute magnitude.  The letter ``p'' appears directly beneath symbols which represent galaxies that are possibly, but unlikely to be, confused (see $\S3.2$).  The solid line represents a linear least squares fit to the filled symbols.}
\end{figure}

\begin{figure}
\plotone{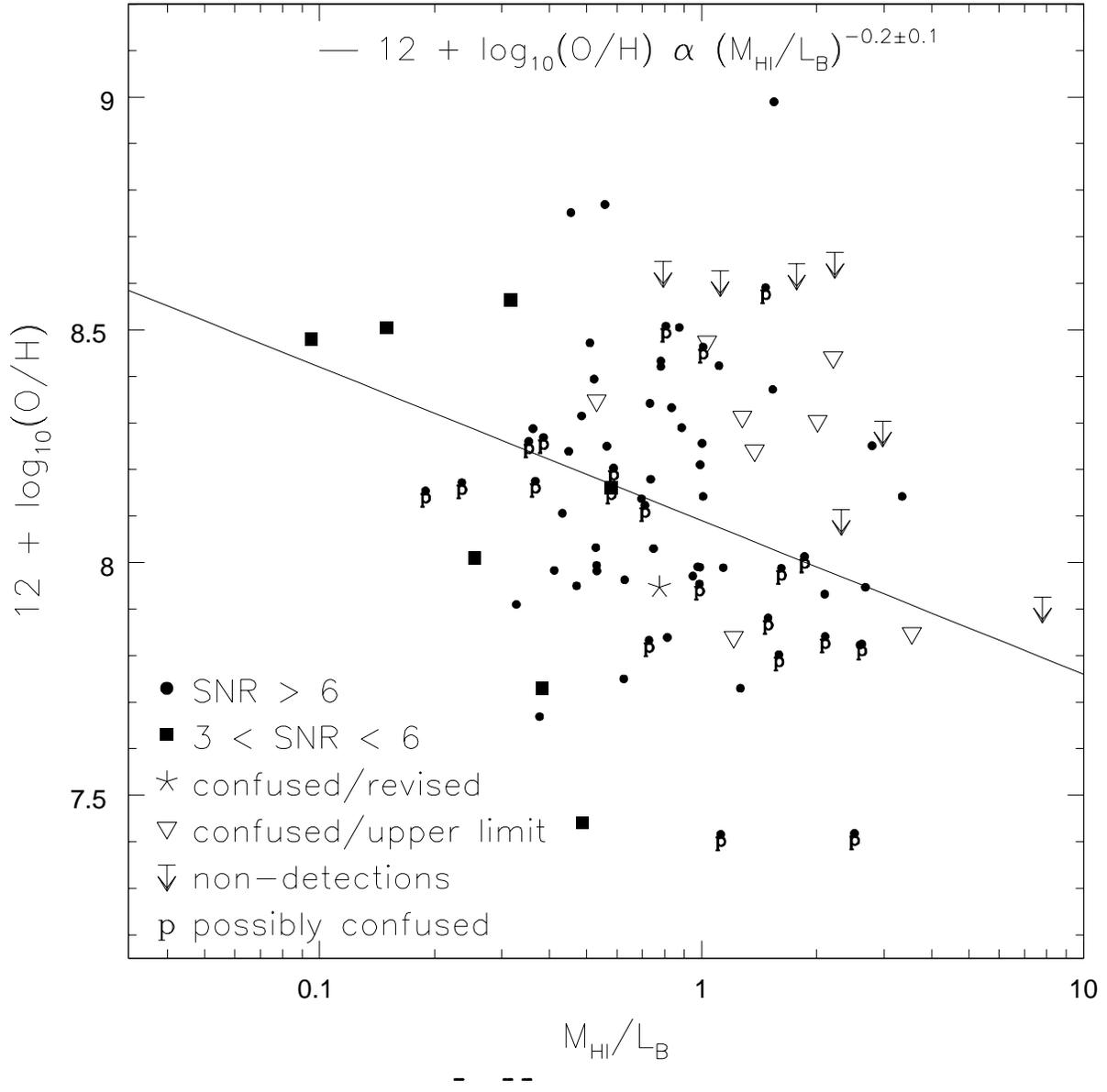}
\figcaption{Metallicity, as given by the relative oxygen abundance, plotted against the ratio of HI mass to blue luminosity.  The letter ``p'' appears directly beneath symbols which represent galaxies that are possibly, but unlikely to be, confused (see $\S3.2$).  The solid line represents a linear least squares fit to the filled symbols.}
\end{figure}

\begin{figure}
\plotone{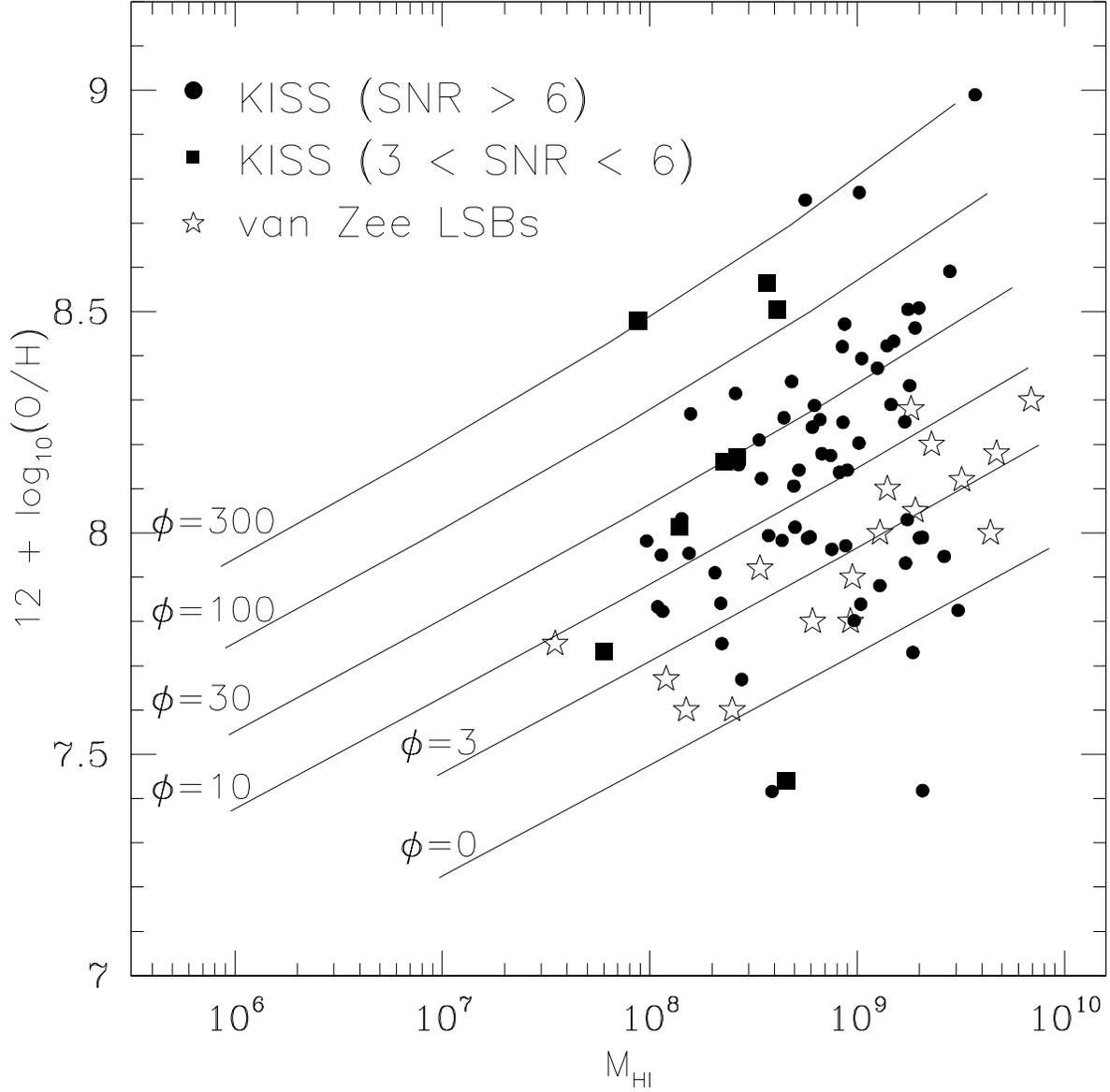}
\figcaption{Metallicity, as given by the oxygen abundance, plotted against the HI gas mass.  Model results from Ferrara \& Tolstoy are over-plotted for lines of constant non-baryonic, dark matter mass to visible mass ratios ($\phi=M_{dark}/M_{stars+gas}$).  Values of $\phi$ are indicated to the left of the corresponding line.  Non-detections and confused sources are omitted.  The high $(M_{HI}/L_B)$ low surface brightness galaxies of van Zee (1997a,b) are represented by open stars, while the KISS galaxies are represented by the filled symbols.}
\end{figure}

\begin{figure}
\plotone{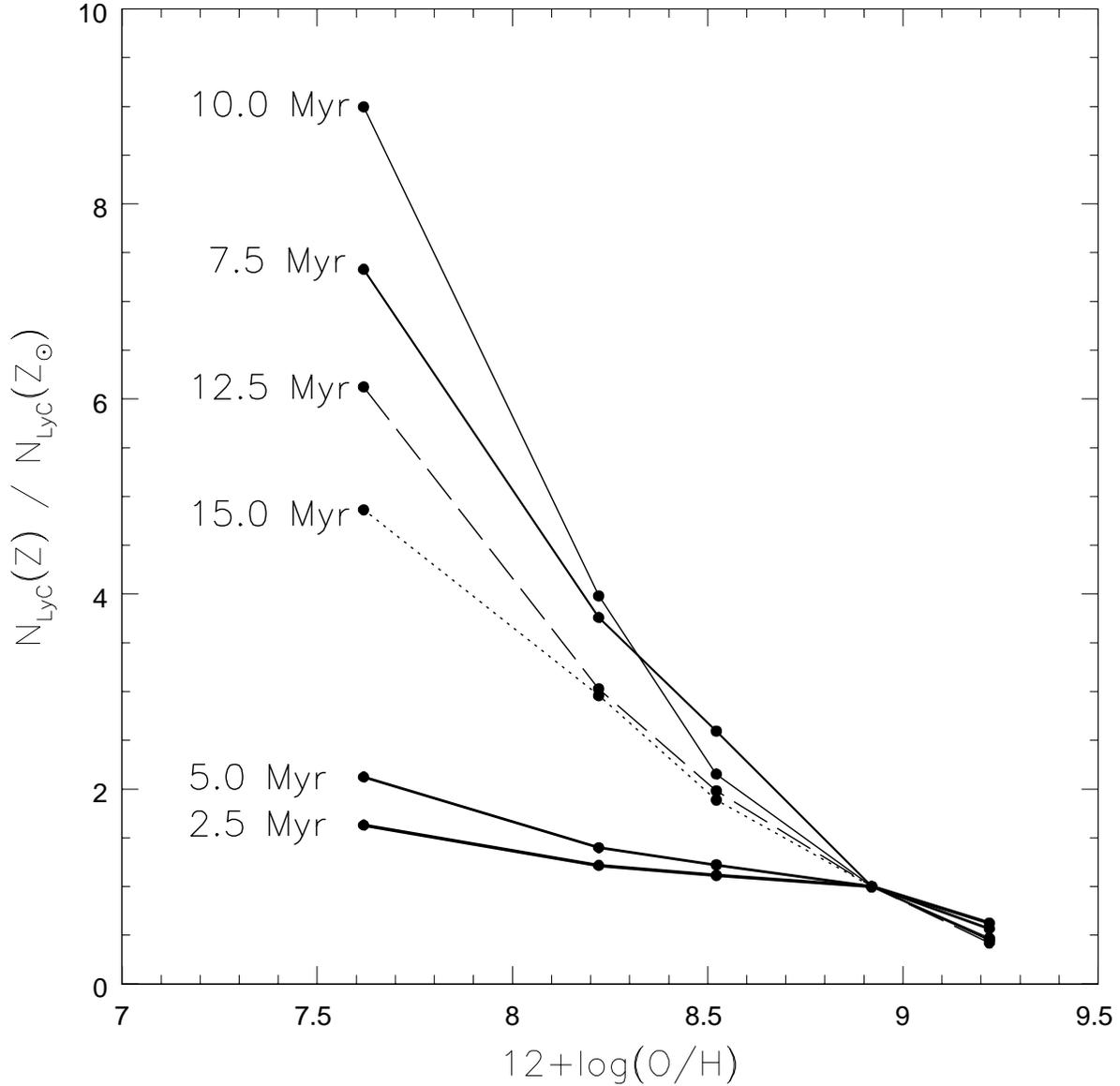}
\figcaption{The number of Lyman continuum photons produced in an instantaneous burst model with metallicity $Z$ normalized by the number produced in a model with solar metallicity, as a function of the metallicity.  The numbers to the left of the curves give the age since the initiation of the star formation event, and the different line thicknesses and types are used to make the curves more distinct.  This range represents the ages during which systems are luminous enough in H$\alpha$ to be observable by KISS, assuming that 0.45 H$\alpha$ photons are produced for every Lyman continuum photon and that the KISS detection limit is nearly zero at $L_{H\alpha} \sim 10^{39} $ergs s$^{-1}$.  For the analyses in this paper, metallicity corrections are based on the ratios on the 15 Myr curve, and errors are based on the range between the highest and lowest possible ratios for a given abundance.  Plot based on the Starburst99 models of Leitherer et al. (1999).}
\end{figure}

\begin{figure}
\plotone{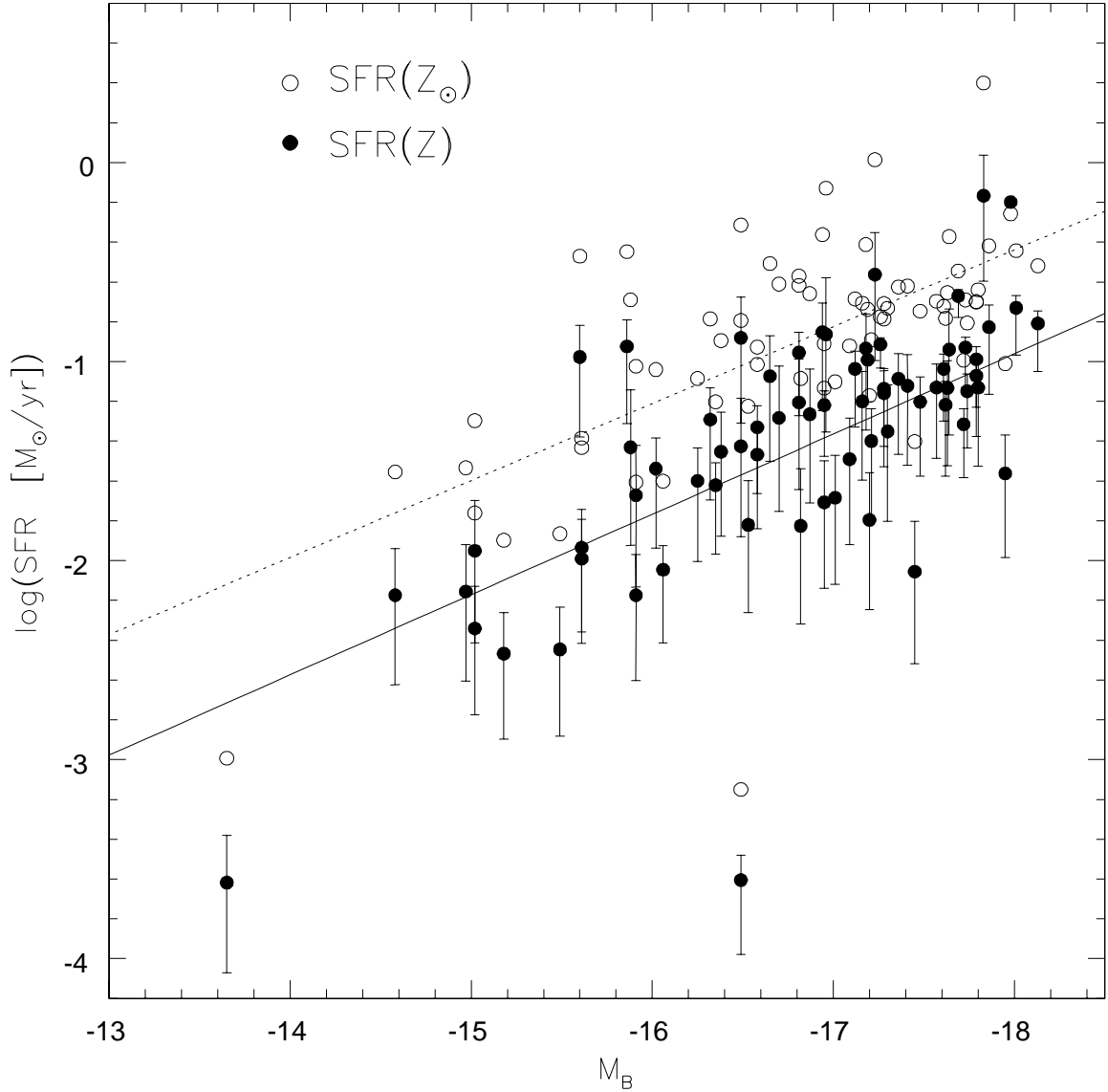}
\figcaption{SFR computed using the H$\alpha$ based Kennicutt (1998) estimator (open circles), and SFR adjusted for metallicity based on the ratios shown in Figure 8 (filled symbols), plotted against the blue absolute magnitudes of the galaxies.  The error bars represent the range of possible SFRs between a burst age of 2.5 Myr to 15.0 Myr.  Least square regressions are shown for both sets of SFRs.}
\end{figure}

\begin{figure}
\plotone{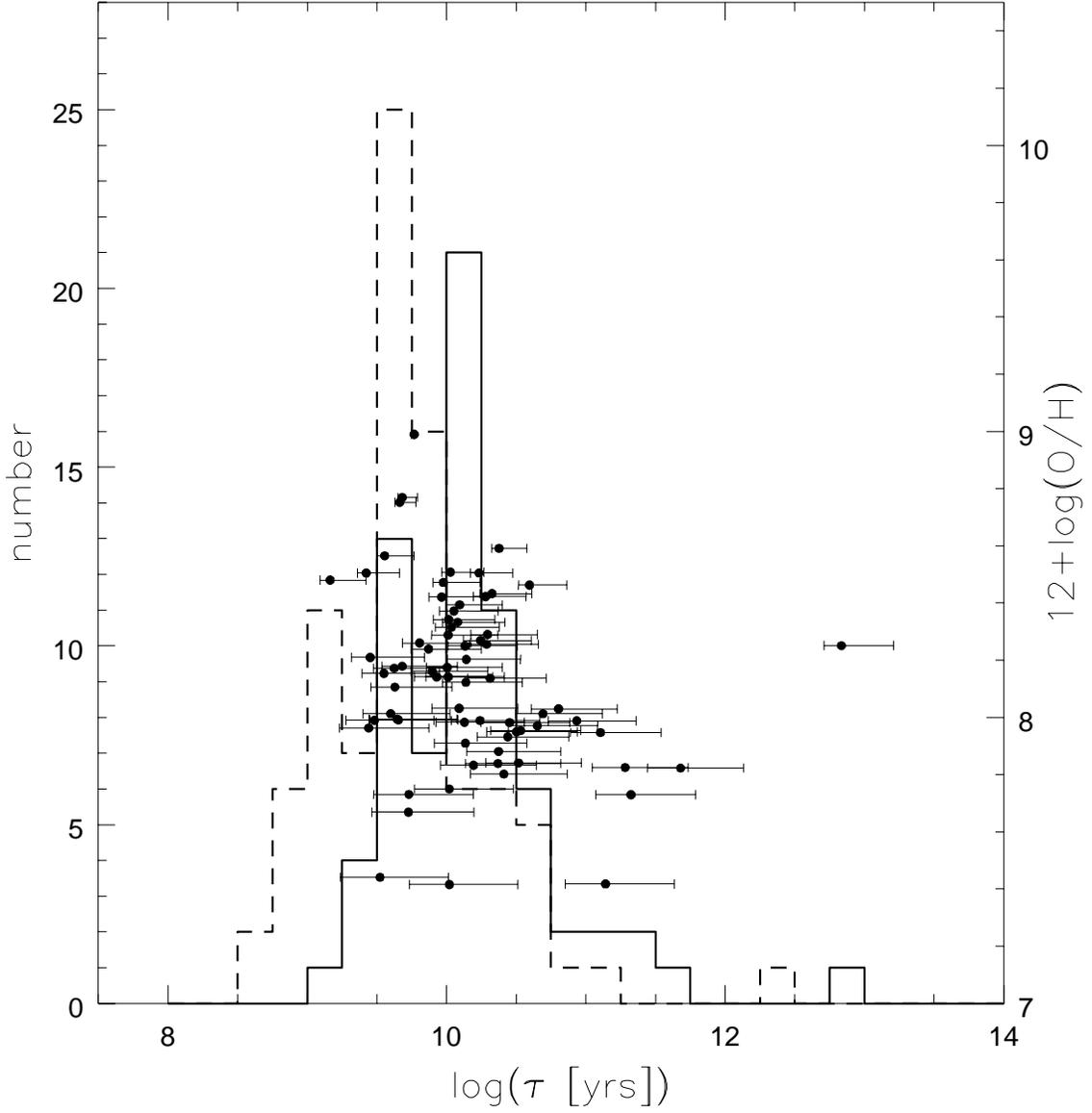}
\figcaption{Histograms of HI gas depletion timescales based a solar metallicity SFR ($\tau(Z_{\odot})$, dotted line) and a SFR adjusted for metallicity ($\tau(Z)$, solid line).  The metallicity given by 12+log(O/H) for $\tau(Z)$ is over-plotted against $\tau$ with error bars indicating the possible variation with age.}
\end{figure}

\begin{figure}
\plotone{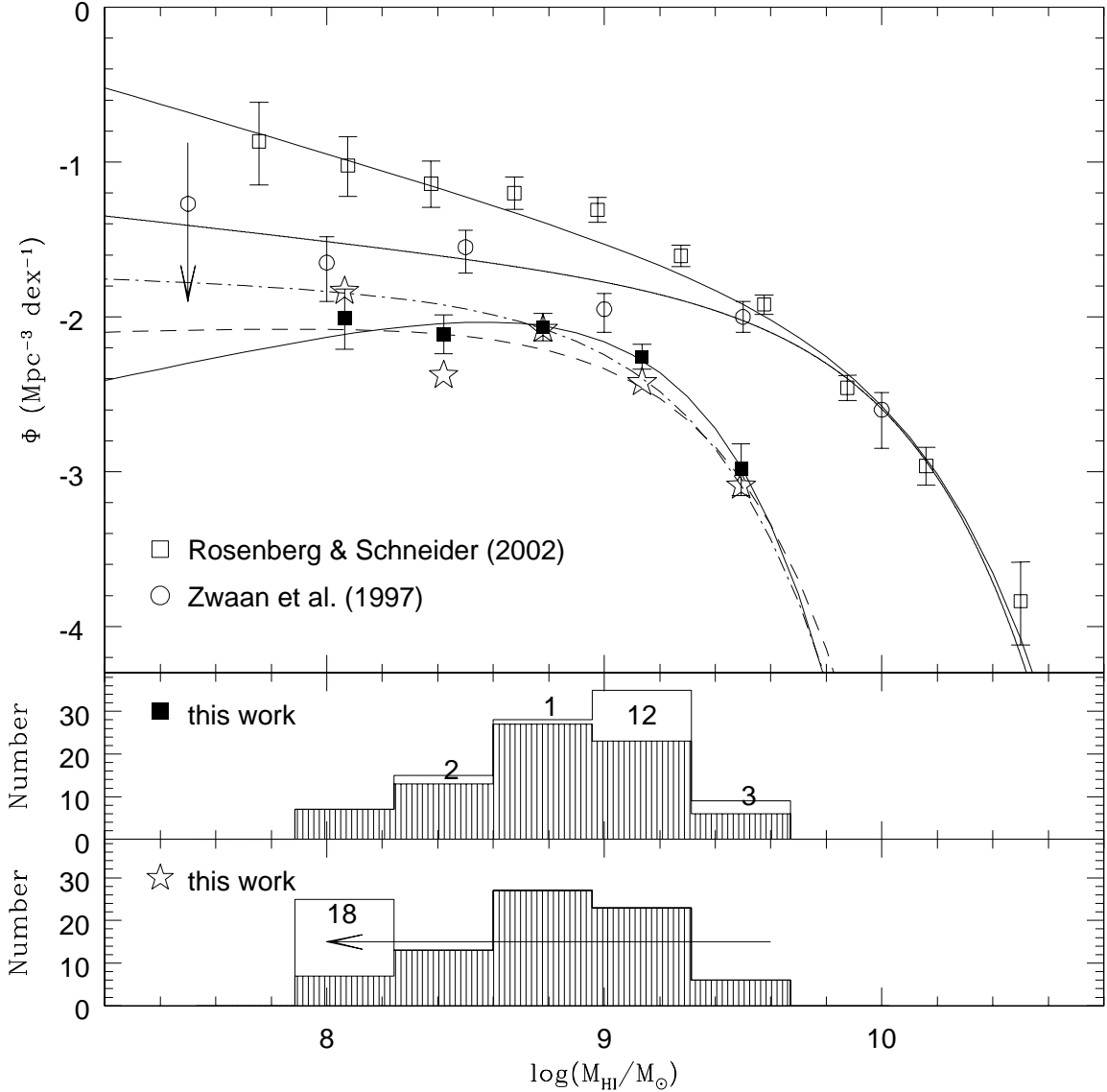}
\figcaption{{\it Top panel:} HI mass functions and Schechter function fits from Rosenberg \& Schneider (2001; open squares/solid curve) and Zwaan et al. (1997; open circles/solid curve), and for the KISS HI sample (solid squares/solid curve and open stars/dashed curve/dot-dashed curve) computed via the $\Sigma$ (1/$V_{max}$) method.  The solid squares represent the HIMF as computed with upper-limit detections placed in bins determined by the value of the upper-limit, whereas the open stars represent the HIMF with all of the upper-limit detections deposited in the lowest mass bin.  Two fits to the open stars are shown: the lower curve (dashed line) is a fit to all the points, while the upper curve (dot-dashed line) shows the fit with the point at the second lowest mass bin (log(M$_{HI}$/M$_{\odot}$)=8.421) removed.  The parameterizations of the Schechter models are: log$(M_{\ast}/M_{\odot}$)=8.96, $\Phi_{\ast}$=0.0087, $\alpha$=-0.60 (solid curve), log$(M_{\ast}/M_{\odot}$)=9.09, $\Phi_{\ast}$=0.0046, $\alpha$=-0.94 (dashed curve), and log$(M_{\ast}/M_{\odot}$)=9.05, $\Phi_{\ast}$=0.0063, $\alpha$=-1.06 (dot-dashed curve).  {\it Middle and lower panels:} Histograms showing the number of galaxies used to compute each of the points in the HIMFs shown in the upper panel.  The middle panel show the distribution for the solid square and the lower panel shows that for the open stars.  Unshaded portions of the histograms represent upper-limit detections while shaded areas show true detections.}
\end{figure}

\clearpage
\renewcommand{\arraystretch}{.6}
\begin{deluxetable}{rcl@{\hspace{1.5mm}}r@{\hspace{1.5mm}}rr@{\hspace{1.5mm}}r@{\hspace{1.5mm}}rcccrrrrrcccr}
\tabletypesize{\tiny}
\rotate
\tablecolumns{20}
\tablewidth{0pt}
\tablecaption{KISS HI Sample\label{table:tab2}}

\tablehead{
\\

&
&
&
&
&
&
&
&
\multicolumn{4}{c}{\scriptsize Optical Data}&
&
\multicolumn{7}{c}{\scriptsize HI Data}
\\\\

\cline{9-11}\cline{13-20}
\\

\colhead{\raisebox{0ex}[0pt]{\scriptsize KISSR}}&
&
&
&
&
&
&
&
&
&
&
&
&
&
&
&
\colhead{RMS}&
\colhead{Order}&
\colhead{Time on}&
\\

&
\colhead{see}&
\multicolumn{3}{c}{\raisebox{-1ex}[0pt]{\scriptsize $\alpha$}}&
\multicolumn{3}{c}{\raisebox{-1ex}[0pt]{\scriptsize $\delta$}}&
\colhead{\raisebox{1ex}[0pt]{B}}&
\colhead{\raisebox{1ex}[0pt]{M$_B$}}&
\colhead{\raisebox{1ex}[0pt]{\bv}}&
&
\colhead{\raisebox{1ex}[0pt]{$v_{helio}$}}&
\colhead{\raisebox{1ex}[0pt]{$\Delta v_{50}$}}&
\colhead{\raisebox{1ex}[0pt]{$\Delta v_{20}$}}&
\colhead{\raisebox{1ex}[0pt]{F.I.}}&
\colhead{of Fit}&
\colhead{of Fit}&
\colhead{Source}&
\colhead{\raisebox{1ex}[0pt]{SNR}}
\\

\colhead{\raisebox{1.5ex}[0pt]{\scriptsize ID}}&
\colhead{\raisebox{1ex}[0pt]{note}}&
\multicolumn{6}{c}{\raisebox{2ex}[0pt]{(J2000)}}&
&
&
&
&
\colhead{(km s$^{-1})$}&
\colhead{(km s$^{-1})$}&
\colhead{(km s$^{-1})$}&
\colhead{(Jy$\cdot$km s$^{-1})$}&
\colhead{(mJy)}&
&
\colhead{(s)}&
\\

\cline{9-11}\cline{13-20}\\

\colhead{(1)}&& 
\multicolumn{3}{c}{(2)} & 
\multicolumn{3}{c}{(3)} & \colhead{(4)} &
\colhead{(5)} & \colhead{(6)} &&\colhead{(7)}& \colhead{(8)} &
\colhead{(9)} & \colhead{(10)} & \colhead{(11)} & \colhead{(12)} &
\colhead{(13)} & \colhead{(14)}}

\startdata  
   1&*&12&15& 6.9&29& 1& 9& 17.35&-17.72&0.53&&\phm{(} 7432\phm{)}&\phm{(}162\phm{)}&\phm{(}184\phm{)}&(0.8123)& 0.7027&5&3000& 36.8\\[0.2ex]
  32&*&12&21&11.4&29&32& 8& 18.64&-16.58&0.61&&\phm{(} 7933\phm{)}&\phm{(}95\phm{)}&\phm{(}182\phm{)}&\phm{(} 0.1894\phm{)}& 0.4843&4&3600& 13.2\\[0.2ex]
  40&*&12&22&23.6&29&26&37& 17.41&-17.80&0.60&&\phm{(} 7827\phm{)}&\phm{(}122\phm{)}&\phm{(}138\phm{)}&\phm{(} 0.2921\phm{)}& 0.4680&4&3000& 23.4\\[0.2ex]
  49&* &12&24&35.2&29&27&32& 17.84&-15.60&0.61&&\phm{(} 8012\phm{)}&\phm{(}103\phm{)}&\phm{(}182\phm{)}&\phm{(}0.3514\phm{)}& 0.6744&6&3300& 16.8\\[0.2ex]
  52& &12&25&25.6&29&44&18& 17.27&-17.99&0.82&&\phm{(} 8057\phm{)}&\phm{(}285\phm{)}&\phm{(}334\phm{)}&\phm{(} 0.5439\phm{)}& 0.5262&4&3000& 24.7\\[0.2ex]
  55&*&12&26&22.0&29&22&12& 19.38&-15.91&0.22&&\phm{(} 8221\phm{)}&\phm{(}112\phm{)}&\phm{(}139\phm{)}&\phm{(} 0.0825\phm{)}& 0.4556&6&2400&  6.8\\[0.2ex]
  57&&12&26&39.3&29&37&58& 17.57&-17.28&0.53&&\phm{(} 6603\phm{)}&\phm{(}140\phm{)}&\phm{(}161\phm{)}&(0.7546)& 0.6966&6&1200& 35.1\\[0.2ex]
  59& &12&27& 8.7&28&57&22& 17.38&-17.73&0.72&&\phm{(} 7636\phm{)}&\phm{(}306\phm{)}&\phm{(}328\phm{)}&(1.1092)& 0.5256&3&2058& 51.0\\[0.2ex]
  61&*&12&27&33.4&29&36& 6& 19.22&-15.61&0.55&&\phm{(} 6560\phm{)}& \phm{(}65\phm{)}& \phm{(}91\phm{)}&\phm{(} 0.0799\phm{)}& 0.5558&2&1200&  6.5\\[0.2ex]
  73&*&12&31&57.2&29&42&46& 15.68&-13.65&0.39&&\phm{(}642\phm{)}& \phm{(}67\phm{)}& \phm{(}83\phm{)}&(7.4010)& 0.8390&5& 600&402.5\\[1.5ex]
  75&*&12&32&48.2&29&23&27& 18.02&-17.18&0.45&&\phm{(} 8021\phm{)}&\phm{(}147\phm{)}&\phm{(}163\phm{)}&\phm{(} 0.6599\phm{)}& 0.6063&2&1200& 36.5\\[0.2ex]
  85&*&12&37&18.5&29&14&54& 19.91&-15.02&0.13&&\phm{(} 6943\phm{)}& \phm{(}57\phm{)}& \phm{(}76\phm{)}&\phm{(} 0.0301\phm{)}& 0.4299&4&1500&  3.5\\[0.2ex]
  91& &12&39&39.6&29&36&34& 17.84&-17.12&0.54&&\phm{(} 7024\phm{)}&\phm{(}189\phm{)}&\phm{(}240\phm{)}&\phm{(} 0.4160\phm{)}& 0.4921&2&2400& 23.8\\[0.2ex]
  96&*&12&40&43.8&29&23& 9& 18.72&-16.81&0.43&&\phm{(} 9199\phm{)}& \phm{(}55\phm{)}&\phm{(}126\phm{)}&\phm{(} 0.5043\phm{)}& 0.5696&4&2418& 30.3\\[0.2ex]
  97&*&12&40&44.3&29&32&57& 18.59&-17.01&0.45&&\phm{(} 9444\phm{)}&\phm{(}184\phm{)}&\phm{(}219\phm{)}&\phm{(} 0.7253\phm{)}& 0.5461&3&2100& 35.6\\[0.2ex]
 105&*&12&42&53.6&29&17&18& 20.29&-14.58&0.50&&\phm{(} 6812\phm{)}& \phm{(}84\phm{)}&\phm{(}111\phm{)}&\phm{(} 0.1179\phm{)}& 0.5167&4&2550&  9.6\\[0.2ex]
 108&*&12&43&55.3&29&22&10& 17.54&-17.41&0.37&&\phm{(} 7082\phm{)}& \phm{(}89\phm{)}& \phm{(}94\phm{)}&\phm{(} 0.1316\phm{)}& 0.3939&4&3798& 14.9\\[0.2ex]
 109& &12&44&16.9&28&47&52& 17.75&-18.13&0.51&&\phm{(}10731\phm{)}&\phm{(}111\phm{)}&\phm{(}116\phm{)}&\phm{(} 0.0877\phm{)}& 0.6504&3&1200&  5.4\\[0.2ex]
 115&* &12&46& 9.1&28&57&30& 17.14&-17.79&0.60&&\phm{(} 7079\phm{)}&\phm{(}148\phm{)}&\phm{(}178\phm{)}&\phm{(} 0.5159\phm{)}& 0.7142&2&1200& 23.5\\[0.2ex]
 116&*&12&46&38.7&29&27&37& 18.40&-17.18&0.43&&\phm{(} 9530\phm{)}&\phm{(}118\phm{)}&\phm{(}148\phm{)}&\phm{(} 0.1344\phm{)}& 0.5897&5&2304&  8.1\\[1.5ex]
 119&*&12&47&24.5&29&12&25& 18.55&-16.35&0.52&&\phm{(} 6993\phm{)}& \phm{(}73\phm{)}&\phm{(}123\phm{)}&\phm{(} 0.3454\phm{)}& 0.6933&4&1314& 18.6\\[0.2ex]
 120&* &12&47&45.7&28&53& 4& 20.75&-14.78&0.49&&(9356)&\nodata&\nodata&\phm{(}$<$ 0.2796\phm{)}& 1.0018&1&1200&\nodata\\[0.2ex]
 125&*&12&48&38.4&29&11&24& 17.20&-17.74&0.44&&\phm{(} 7146\phm{)}&\phm{(}154\phm{)}&\phm{(}203\phm{)}&\phm{(} 0.7364\phm{)}& 0.6189&2&1200& 36.5\\[0.2ex]
 133&&12&51& 6.6&29&11&48& 17.85&-16.95&0.54&&\phm{(} 6723\phm{)}& \phm{(}40\phm{)}& \phm{(}55\phm{)}&\phm{(} 0.0482\phm{)}& 0.4996&4&2100&  5.8\\[0.2ex]
 142& &12&53&49.2&28&56&33& 17.20&-17.75&0.88&&(7143)&\nodata&\nodata&\phm{(}$<$ 0.8357\phm{)}& 0.8809&1& 600&\nodata\\[0.2ex]
 146$^1$&*&12&54&36.8&28&55&48& 16.70&-15.81&0.09&&(2327)&\nodata&\nodata&\nodata& \nodata&\nodata&\nodata&\nodata\\[0.2ex]
 148$^1$&*&12&54&45.3&28&55&29& 17.06&-15.50&0.15&&(2393)&\nodata&\nodata&\nodata& \nodata&\nodata&\nodata&\nodata\\[0.2ex]
 156&*&12&57&43.6&29& 0&11& 17.94&-16.95&0.25&&\phm{(} 6962\phm{)}&\phm{(}127\phm{)}&\phm{(}181\phm{)}&\phm{(} 0.4507\phm{)}& 0.6029&6&1200& 24.2\\[0.2ex]
 170&&13& 0&31.3&28&57& 1& 17.72&-17.19&0.57&&\phm{(} 7005\phm{)}&\phm{(}243\phm{)}&\phm{(}270\phm{)}&(0.1746)& 0.8162&4&1200&  5.8\\[0.2ex]
 171&* &13& 0&37.2&28&39&50& 17.96&-16.99&0.73&&(7096)&\nodata&\nodata&\phm{(}$<$ 0.5279\phm{)}& 0.6483&5&1200&\nodata\\[1.5ex]
 182&*&13& 2&25.7&28&51&28& 17.19&-17.63&0.44&&\phm{(} 6701\phm{)}&\phm{(}154\phm{)}&\phm{(}194\phm{)}&\phm{(} 0.5587\phm{)}& 0.6570&3&1800& 25.6\\[0.2ex]
 187&*&13& 4&16.8&28&51& 2& 18.83&-16.25&0.50&&\phm{(} 7584\phm{)}&\phm{(}195\phm{)}&\phm{(}257\phm{)}&\phm{(} 0.1497\phm{)}& 0.4612&4&2304&  8.6\\[0.2ex]
 191& &13& 4&38.8&28&58&21& 17.13&-17.95&0.38&&\phm{(} 7581\phm{)}&\phm{(}127\phm{)}&\phm{(}195\phm{)}&\phm{(} 0.7545\phm{)}& 0.9141&3&1200& 24.9\\[0.2ex]
 192&*&13& 5& 6.5&28&38&28& 18.55&-15.86&0.88&&\phm{(} 5496\phm{)}&\phm{(}179\phm{)}&\phm{(}208\phm{)}&\phm{(} 0.2683\phm{)}& 0.5010&1&2400& 16.6\\[0.2ex]
 193&*&13& 5&12.3&29&14& 9& 18.90&-16.02&0.51&&\phm{(} 6965\phm{)}&\phm{(}224\phm{)}&\phm{(}262\phm{)}&\phm{(} 0.1143\phm{)}& 0.5564&5&2100&  5.5\\[0.2ex]
 194&*&13& 5&15.5&28&37&35& 17.50&-17.28&0.51&&\phm{(} 6601\phm{)}&\phm{(}167\phm{)}&\phm{(}202\phm{)}&\phm{(} 0.2485\phm{)}& 0.5125&5&2100& 14.9\\[0.2ex]
 195& &13& 5&32.8&29& 0&41& 16.34&-17.97&0.67&&\phm{(} 5247\phm{)}&\phm{(}286\phm{)}&\phm{(}317\phm{)}&(0.6664)& 0.6105&6&2100& 26.6\\[0.2ex]
 205&*&13& 7&23.0&29&24& 4& 17.18&-17.09&0.39&&\phm{(} 5254\phm{)}& \phm{(}75\phm{)}&\phm{(}173\phm{)}&\phm{(} 0.3868\phm{)}& 0.4888&4&2400& 22.4\\[0.2ex]
 207& &13& 8& 4.0&28&59&54& 17.20&-17.83&0.44&&\phm{(} 7230\phm{)}&\phm{(}192\phm{)}&\phm{(}243\phm{)}&(0.8649)& 0.6201&5&1200& 39.0\\[0.2ex]
 210& &13& 8&15.2&29& 1&22& 18.07&-16.57&0.35&&\phm{(} 6155\phm{)}& \phm{(}95\phm{)}&\phm{(}140\phm{)}&\phm{(} 0.4788\phm{)}& 0.6485&4&1200& 26.3\\[1.5ex]
 215&*&13& 8&54.9&29&32&39& 17.02&-17.83&0.60&&(6820)&\nodata&\nodata&\phm{(}$<$ 0.7169\phm{)}& 0.7557&2&1146&\nodata\\[0.2ex]
 217&* &13& 9& 7.3&28&40& 6& 16.82&-17.61&0.47&&\phm{(} 5621\phm{)}&\phm{(}123\phm{)}&\phm{(}238\phm{)}&\phm{(} 0.6731\phm{)}& 0.5328&3&2100& 36.2\\[0.2ex]
 236& &13&13&35.3&29& 7&35& 17.19&-17.36&0.96&&\phm{(} 5999\phm{)}&\phm{(}311\phm{)}&\phm{(}332\phm{)}&(0.6000)& 0.5062&3&2316& 28.4\\[0.2ex]
 238&*&13&14&37.6&29&19& 4& 17.47&-17.28&0.74&&\phm{(} 6680\phm{)}& \phm{(}43\phm{)}& \phm{(}54\phm{)}&\phm{(} 0.2241\phm{)}& 0.3785&3& 900& 31.6\\[0.2ex]
 245& &13&16&28.0&29&25&11& 17.71&-16.35&0.49&&\phm{(} 4810\phm{)}& \phm{(}85\phm{)}&\phm{(}197\phm{)}&\phm{(} 0.2768\phm{)}& 0.5527&2&1800& 15.0\\[0.2ex]
 252&*&13&17&56.7&28&42&41& 17.96&-17.79&0.59&&\phm{(}10400\phm{)}&\phm{(}185\phm{)}&\phm{(}221\phm{)}&\phm{(} 0.4032\phm{)}& 0.7260&5&1800& 16.4\\[0.2ex]
 256&*&13&18&22.4&28&49&31& 18.83&-16.95&0.42&&\phm{(}10433\phm{)}&(323)&(323)&\phm{(}$<$ 0.2677\phm{)}& 0.5418&6&1800& 13.0\\[0.2ex]
 257&*&13&18&25.8&28&47&40& 18.40&-17.37&0.47&&\phm{(}10445\phm{)}&(177)&(218)&\phm{(}$<$ 0.1639\phm{)}& 0.4826&5&3186& 10.0\\[0.2ex]
 265&*&13&19&51.5&29&25&31& 18.64&-16.94&0.47&&\phm{(} 9496\phm{)}&(176)&(203)&\phm{(}$<$ 0.5512\phm{)}& 0.5712&2&2112& 28.5\\[0.2ex]
 271& &13&21&40.8&28&52&59& 17.47&-17.36&0.69&&\phm{(} 6723\phm{)}&\phm{(}164\phm{)}&\phm{(}240\phm{)}&\phm{(} 0.3253\phm{)}& 0.5875&2&1500& 15.5\\[1.5ex]
 272&*&13&21&45.1&29&27&51& 18.25&-17.20&0.36&&\phm{(} 8970\phm{)}&\phm{(}153\phm{)}&\phm{(}189\phm{)}&\phm{(} 0.9341\phm{)}& 0.5383&5&1158& 51.7\\[0.2ex]
 278& &13&23&37.7&29&17&17& 18.23&-15.49&0.41&&\phm{(} 4053\phm{)}& \phm{(}53\phm{)}&\phm{(}108\phm{)}&\phm{(} 0.1655\phm{)}& 0.3697&5&3600& 18.7\\[0.2ex]
 280&*&13&24& 8.7&29&11& 3& 18.46&-15.91&0.53&&\phm{(} 5442\phm{)}&\phm{(}195\phm{)}&\phm{(}246\phm{)}&\phm{(} 0.4708\phm{)}& 0.4949&5&2400& 26.3\\[0.2ex]
 286&*&13&26&25.1&29&10&31& 17.74&-16.49&0.56&&\phm{(} 5198\phm{)}&\phm{(}211\phm{)}&\phm{(}277\phm{)}&\phm{(} 0.5445\phm{)}& 0.4212&6&2400& 33.3\\[0.2ex]
 299&*&13&29&56.5&29&46&19& 17.99&-17.16&0.26&&\phm{(} 7798\phm{)}&\phm{(}114\phm{)}&\phm{(}133\phm{)}&\phm{(} 0.1061\phm{)}& 0.6767&4&1200&  6.0\\[0.2ex]
 305& &13&32&36.1&29& 6&45& 17.83&-17.98&0.66&&\phm{(}10633\phm{)}&\phm{(}309\phm{)}&\phm{(}320\phm{)}&\phm{(} 0.8133\phm{)}& 0.9660&4&1500& 20.6\\[0.2ex]
 310&*&13&33&45.3&28&45&12& 18.86&-16.96&0.60&&\phm{(}10619\phm{)}&\phm{(}168\phm{)}&\phm{(}174\phm{)}&\phm{(} 0.0987\phm{)}& 0.7325&1&1200&  4.4\\[0.2ex]
 311& &13&33&55.8&29&21&51& 18.90&-16.70&0.62&&\phm{(} 9640\phm{)}& \phm{(}91\phm{)}&\phm{(}118\phm{)}&\phm{(} 0.0741\phm{)}& 0.4907&5&2322&  6.1\\[0.2ex]
 314&* &13&35&35.6&29&13& 0& 15.25&-15.18&0.34&&\phm{(}  858\phm{)}& \phm{(}91\phm{)}&\phm{(}130\phm{)}&(2.3800)& 1.0232&4& 600& 81.5\\[0.2ex]
 356&*&13&49&27.8&29&42& 5& 17.82&-17.94&0.84&&\phm{(}10486\phm{)}& (34)& (40)&\phm{(} 0.1069\phm{)}& 0.7146&1&1200& 10.3\\[1.5ex]
\\\\\\\\\\\\\\\\\\\\\\\\\\\\\\\\\\\\\\\\\\\
 368& &13&51& 1.2&29&17& 5& 18.12&-17.69&0.59&&\phm{(}10735\phm{)}&\phm{(}127\phm{)}&\phm{(}155\phm{)}&\phm{(} 0.2217\phm{)}& 0.6653&2&2280& 11.7\\[0.2ex]
 386& &13&54&25.8&29&33& 0& 17.32&-17.62&0.65&&\phm{(} 7221\phm{)}&\phm{(}246\phm{)}&\phm{(}268\phm{)}&\phm{(} 0.2924\phm{)}& 0.4842&3&2700& 16.2\\[0.2ex]
 396&*&13&57&10.0&29&13&10& 17.49&-14.97&0.34&&\phm{(} 2258\phm{)}& \phm{(}47\phm{)}&\phm{(}141\phm{)}&\phm{(} 0.5139\phm{)}& 0.6499&3&1200& 28.4\\[0.2ex]
 398&* &13&57&18.1&28&42&19& 18.81&-16.93&0.86&&(10387)&\nodata&\nodata&\phm{(}$<$ 0.4662\phm{)}& 0.5895&3&1200&\nodata\\[0.2ex]
 401$^2$&*&13&57&21.8&28&47&12& 18.43&-17.45&0.75&&(10905)&\nodata&\nodata&\nodata&\nodata&\nodata&\nodata&\nodata\\[0.2ex]
 404&*&13&58&19.3&29&28&24& 19.34&-16.19&0.53&&\phm{(} 9387\phm{)}&(164)&(233)&\phm{(}$<$ 0.4508\phm{)}& 0.5272&4&2046& 24.3\\[0.2ex]
 405&*&13&58&20.0&29&27&58& 18.09&-17.45&0.38&&\phm{(} 9377\phm{)}&(154)&(197)&\phm{(}$<$ 0.4847\phm{)}& 0.5818&4&1770& 25.6\\[0.2ex]
 407&*&13&58&36.1&29&23&20& 16.78&-17.66&0.49&&\phm{(} 5679\phm{)}& \phm{(}56\phm{)}& \phm{(}94\phm{)}&\phm{(} 2.0290\phm{)}& 0.8225&5& 600& 92.0\\[0.2ex]
 460&*&14& 8&18.8&29& 1& 0& 17.11&-17.87&0.86&&( 7323)&\nodata&\nodata&$<$(0.8385)& 0.8838&5&1500&\nodata\\[0.2ex]
 465& &14& 9&37.5&29& 5&35& 18.32&-17.17&0.59&&\phm{(} 9223\phm{)}&\phm{(}123\phm{)}&\phm{(}186\phm{)}&\phm{(} 0.2720\phm{)}& 0.5030&6&2400& 17.2\\[1.5ex]
 471&*&14&10&21.1&29&38&18& 18.20&-16.82&0.15&&\phm{(}10784\phm{)}&\phm{(}191\phm{)}&\phm{(}218\phm{)}&\phm{(} 0.4288\phm{)}& 0.4246&5&2700& 29.8\\[0.2ex]
 505& &14&16&55.4&29&29&11& 16.74&-16.53&0.40&&\phm{(} 3307\phm{)}&\phm{(} 74\phm{)}&\phm{(}100\phm{)}&\phm{(} 0.4519\phm{)}& 0.7478&4&1200& 24.5\\[0.2ex]
 507&*&14&17&24.7&29&41&11& 17.71&-17.59&0.74&&\phm{(} 8525\phm{)}&\phm{(}155\phm{)}&\phm{(}174\phm{)}&\phm{(} 0.1310\phm{)}& 0.4156&3&3516& 10.7\\[0.2ex]
 508&*&14&17&27.4&29& 8& 0& 18.89&-16.94&0.47&&\phm{(}10837\phm{)}&\phm{(}160\phm{)}&\phm{(}166\phm{)}&\phm{(} 0.1421\phm{)}& 0.3758&5&3000& 12.8\\[0.2ex]
 515&*&14&19&10.5&28&55&35& 18.12&-17.73&0.74&&\phm{(}10999\phm{)}&\phm{(}213\phm{)}&\phm{(}244\phm{)}&\phm{(} 0.5723\phm{)}& 0.4090&6&2400& 38.8\\[0.2ex]
 518& &14&19&52.8&29&30&34& 17.46&-18.00&0.52&&( 9149)&\nodata&\nodata&\phm{(}$<$ 0.5603\phm{)}& 0.5906&1&1800&\nodata\\[0.2ex]
 528&*&14&22& 4.2&28&52& 0& 18.96&-16.87&0.47&&\phm{(}10791\phm{)}&\phm{(}141\phm{)}&\phm{(}177\phm{)}&\phm{(} 0.2706\phm{)}& 0.4876&2&2400& 17.4\\[0.2ex]
 541&*&14&23&58.3&29&49&33& 16.85&-16.26&0.50&&\phm{(} 3084\phm{)}&\phm{(}150\phm{)}&\phm{(}171\phm{)}&(0.4064)& 0.3987&3&3000& 34.1\\[0.2ex]
 544& &14&24&24.7&28&44&48& 17.55&-17.86&0.57&&\phm{(} 8921\phm{)}&\phm{(}238\phm{)}&\phm{(}249\phm{)}&(0.5276)& 0.4671&4&2034& 31.2\\[0.2ex]
 561&*&14&29&53.6&29&20&10& 17.86&-15.61&0.31&&\phm{(} 3793\phm{)}&\phm{(}106\phm{)}&\phm{(}127\phm{)}&\phm{(} 0.8864\phm{)}& 0.7931&2&1200& 38.8\\[1.5ex]
 564& &14&45&19.4&29&18&52& 18.06&-17.57&0.58&&\phm{(} 9821\phm{)}&\phm{(}170\phm{)}&\phm{(}176\phm{)}&\phm{(} 0.3713\phm{)}& 0.5510&2&3600& 22.1\\[0.2ex]
 572&*&14&46&48.2&29&25&17& 17.50&-16.06&0.41&&\phm{(} 3743\phm{)}& \phm{(}82\phm{)}&\phm{(}135\phm{)}&\phm{(} 0.2673\phm{)}& 0.5600&5&2400& 17.6\\[0.2ex]
 590&* &14&55& 1.2&29&33&30& 19.28&-16.28&0.53&&( 9251)&\nodata&\nodata&\phm{(}$<$ 0.4248\phm{)}& 0.7360&4&1200&\nodata\\[0.2ex]
 666&*&15&15&42.5&29& 1&39& 19.83&-15.88&0.41&&\phm{(} 9996\phm{)}& \phm{(}58\phm{)}& \phm{(}91\phm{)}&\phm{(} 0.0932\phm{)}& 0.3724&5&2700& 12.1\\[0.2ex]
 675&*&15&17&17.4&29&24&24& 16.83&-17.45&0.45&&\phm{(} 5260\phm{)}& \phm{(}64\phm{)}& \phm{(}99\phm{)}&(1.3620)& 1.7605&2& 600& 34.1\\[0.2ex]
 678& *&15&17&47.7&29&37&55& 18.85&-16.81&0.48&&\phm{(} 9922\phm{)}&\phm{(}173\phm{)}&\phm{(}179\phm{)}&\phm{(} 0.3094\phm{)}& 0.4545&4&3300& 22.3\\[0.2ex]
 742&*&15&26&22.9&29&11&49& 18.92&-16.85&0.32&&(10261)&\nodata&\nodata&\phm{(}$<$ 0.4497\phm{)}& 0.5912&4&1746&\nodata\\[0.2ex]
 756&*&15&27&34.6&29&17&18& 18.19&-17.72&0.58&&(10957)&\nodata&\nodata&\phm{(}$<$ 0.6740\phm{)}& 0.7105&4&1200&\nodata\\[0.2ex]
 757&&15&27&38.4&29&41& 6& 18.24&-17.64&0.34&&\phm{(}10845\phm{)}&\phm{(}163\phm{)}&\phm{(}210\phm{)}&\phm{(} 0.4139\phm{)}& 0.3988&4&3852& 30.5\\[0.2ex]
 785&*&15&31& 9.8&28&51&32& 19.45&-16.49&0.43&&\phm{(}10688\phm{)}&\phm{(}124\phm{)}&\phm{(}169\phm{)}&\phm{(} 0.2064\phm{)}& 0.4516&5&2814& 15.0\\[1.5ex]
 803& &15&33&30.3&29&37& 5& 18.60&-17.30&0.37&&\phm{(}10440\phm{)}&\phm{(}174\phm{)}&\phm{(}190\phm{)}&\phm{(} 0.2318\phm{)}& 0.4396&7&2622& 16.6\\[0.2ex]
 830&*&15&38& 3.0&29&33&20& 17.86&-18.01&0.29&&\phm{(}10647\phm{)}&\phm{(}133\phm{)}&\phm{(}163\phm{)}&\phm{(} 0.4295\phm{)}& 0.4985&5&2922& 28.4\\[0.2ex]
 856&*&15&46&45.5&29&52& 9& 20.94&-14.43&0.66&&( 8364)&\nodata&\nodata&\phm{(}$<$ 0.1726\phm{)}& 0.7326&2&1200&\nodata\\[0.2ex]
 882& &15&52&48.6&28&55&59& 18.53&-17.26&0.67&&\phm{(}10055\phm{)}&\phm{(}156\phm{)}&\phm{(}175\phm{)}&\phm{(} 0.1346\phm{)}& 0.4983&7&3600&  8.9\\[0.2ex]
 891& &15&53&39.8&28&49&28& 18.32&-17.48&0.54&&\phm{(}10060\phm{)}&\phm{(}116\phm{)}&\phm{(}140\phm{)}&\phm{(} 0.2016\phm{)}& 0.4368&1&3210& 15.5\\[0.2ex]
 956& &16& 2& 5.2&29&43&37& 17.39&-16.58&0.44&&\phm{(} 4310\phm{)}&\phm{(}159\phm{)}&\phm{(}199\phm{)}&\phm{(} 0.8338\phm{)}& 0.5645&6&2370& 45.3\\[0.2ex]
 979& &16& 8&34.1&29&45&30& 18.51&-17.21&0.54&&\phm{(} 9910\phm{)}&\phm{(}118\phm{)}&\phm{(}176\phm{)}&\phm{(} 0.2027\phm{)}& 0.4654&3&3300& 13.9\\[0.2ex]
 998& &16&12&23.0&29& 9&41& 19.06&-16.75&0.82&&\phm{(}10240\phm{)}& \phm{(}39\phm{)}& \phm{(}41\phm{)}&\phm{(} 0.0267\phm{)}& 0.6342&2&1200&  2.8\\[0.2ex]
 999&*&16&12&24.4&29&24&19& 17.72&-17.99&1.04&&( 9737)&\nodata&\nodata&\phm{(}$<$ 0.6031\phm{)}& 0.6357&5&1176&\nodata\\[0.2ex]
1005& &16&14& 6.0&29&23&42& 18.54&-17.26&0.74&&\phm{(}10178\phm{)}&\phm{(}130\phm{)}&\phm{(}152\phm{)}&\phm{(} 0.1011\phm{)}& 0.3463&5&2958& 10.4\\[1.5ex]
1011& &16&15&46.5&29&52&53& 16.92&-15.86&0.56&&\phm{(} 2517\phm{)}&\phm{(}155\phm{)}&\phm{(}176\phm{)}&\phm{(} 2.0540\phm{)}& 0.9500&5& 900& 71.0\\[0.2ex]
1013&*&16&16&39.0&29& 3&33& 17.79&-17.34&0.41&&\phm{(} 7475\phm{)}&(146)&(195)&\phm{(} 0.4369\phm{)}& 0.5425&6&1200& 37.3\\[0.2ex]
1014& &16&17&13.7&29&49&34& 19.41&-16.32&0.36&&\phm{(}10009\phm{)}&\phm{(}132\phm{)}&\phm{(}149\phm{)}&\phm{(} 0.1295\phm{)}& 0.5520&4&1506&  9.2\\[0.2ex]
1021&*&16&19& 2.6&29&10&22& 17.85&-15.02&0.32&&\phm{(} 2541\phm{)}&\phm{(}110\phm{)}&\phm{(}128\phm{)}&\phm{(} 0.5065\phm{)}& 0.5303&7&2100& 36.7\\[0.2ex]
1048&* &16&33&47.6&28&59& 5& 14.44&-16.49&0.48&&\phm{(}  965\phm{)}&\phm{(}130\phm{)}&\phm{(}158\phm{)}&(17.7600)& 0.9387&3& 600&651.9\\[0.2ex]
1091& &16&50& 2.7&29&25&55& 19.41&-16.38&0.67&&\phm{(} 9653\phm{)}& \phm{(}60\phm{)}& \phm{(}69\phm{)}&\phm{(} 0.0356\phm{)}& 0.4894&7&1734&  3.7\\[0.2ex]
1112& &16&55&47.9&29&49& 1& 19.24&-16.86&0.51&&\phm{(}10669\phm{)}&\phm{(}144\phm{)}&\phm{(}199\phm{)}&\phm{(} 0.5783\phm{)}& 0.4199&5&2058& 38.7\\[0.2ex]
1121& &16&57&30.2&29& 1& 8& 18.61&-17.23&0.40&&\phm{(} 9530\phm{)}&\phm{(}151\phm{)}&\phm{(}189\phm{)}&\phm{(} 0.1985\phm{)}& 0.4402&7&3210& 14.4\\[0.2ex]
1123& &16&58&52.8&29& 4&15& 19.38&-16.65&0.82&&\phm{(}10439\phm{)}& \phm{(}77\phm{)}& \phm{(}99\phm{)}&\phm{(} 0.0821\phm{)}& 0.5214&6&3300&  7.2\\[0.2ex]
\hline
\multicolumn{20}{l}{\tiny $^1$ 21-cm emission dominated by the nearby galaxy U8033.  Observed quantities for this severely confused region are given in Table 2.}\\[0.2ex]
\multicolumn{20}{l}{\tiny $^2$ 21-cm emission probably dominated by KISSR 399 and 401.  Observed quantities for this severely confused region are given in Table 2.}\\[0.2ex]
\hline

\enddata

\end{deluxetable}


\clearpage
\renewcommand{\baselinestretch}{.9}
\renewcommand{\arraystretch}{.8}

\begin{deluxetable}{rcp{4in}}
\scriptsize
\tablecolumns{3}
\tablewidth{0pt}
\tablecaption{Notes on Individual Galaxies}
\tablehead{
\colhead{KISSR}&\colhead{Other Names}&\colhead{Notes}\\
\colhead{ID}&&\\}
\startdata

   1&\nodata&Possibly, but not likely, confused.  1\farcm2 NNE from faint LSB with no published redshift.  Noisy baseline due to solar interference.\\
  32&\nodata&  Noisy baseline due to solar interference.\\
  40&\nodata&Possibly, but not likely, confused.  3\farcm3 W from KISSR 41 ($cz$ = 7910 km/s) which is near first null of beam.  Noisy baseline due to solar interference.\\

  49&CG 177&Noisy baseline due to solar interference.\\

  55&\nodata& 5\farcm7 WSW from faint edge-on galaxy in sidelobe with no published redshift.  Unlikely to be confused.\\

  61&\nodata& 3\farcm5 WSW from faint galaxy with no published redshift.  3\farcm8 N from second faint galaxy with no published redshift.  Unlikely to be confused.\\
 
  73&\nodata& KISSR 73 is HII region in LSB dwarf galaxy.\\
 
  75&\nodata& 5\farcm9 ESE from faint galaxy in sidelobe with no published redshift.  Unlikely to be confused.\\

  85&\nodata&5\farcm2 WNW from faint galaxy with no published redshift in sidelobe.  Unlikely to be confused.\\

  96&CG 189& 5\farcm5 SSW from UGC 7836 (Scd, m$_B$ = 14.83, $cz$ = 9353 km s$^{-1}$) in sidelobe.  However, no signal detected at 9353 km/s.  Unlikely to be confused.\\

  97&\nodata& 5\farcm7 NNW from UGC 7836 (Scd, m$_B$ = 14.83, $cz$ = 9353 km $s^{-1}$) in sidelobe.  However, no signal detected at 9353 km/s.  Unlikely to be confused.\\
 105&\nodata&Possibly, but not likely, confused.  2\farcm5 SW from faint LSB galaxy with no published redshift.\\

 108&CG 192&Possibly, but not likely, confused.  0\farcm6 N from very faint galaxy with no published redshift.\\ \\

 115&CG 194&  Noisy baseline due to solar interference.\\

 116&CG 195, Was 63& 4\farcm7 N from bright galaxy near first null.  Unlikely to be confused.\\

 119&\nodata&3\farcm0 ENE from bright edge-on spiral ($cz$ = 6895 km s$^{-1}$) near first null.  Unlikely to be confused.\\

 120&&Non-detection, noisy baseline due to solar interference.\\

 125&\nodata&4\farcm6 SSE from galaxy near first null with no published redshift.  Unlikely to be confused\\
\\\\\\\\\\\\\\\\\\\\\\\\\\\\\\\\\\\\\\\\\\\\\\\\\\\\\\\\\\\\\\\\\\\\\\\\\\\\\\\\\\\\\\\\\\\\\\\\\\\\\\\\\\\\\\\
 142&\nodata& Non-detection, dE galaxy optical spectrum.\\

 146&\nodata& Confused. Signal swamped by UGC 8033 (KISSR 147;$cz$ = 2485 \kms), 1\arcmin away.  Observed quantities for this region are $v_{helio}$=2501 \kms, $\Delta v_{50}$=306 \kms, $\Delta v_{20}$=364 \kms, F.I. = 2.963 Jy \kms, RMS = 2.6428 mJy, Order = 2, Time = 600 s, SNR = 24.8.\\\

 148&\nodata& Confused. Signal swamped by UGC 8033 (KISSR 147;$cz$ = 2485 km/s) 1\farcm3 away.  Observed quantities for this region are given in the notes to KISSR 146. \\

 156&&Emission region offset from center of galaxy.\\ 

 171&\nodata& Non-detection.\\

 182&CG 963,&Possibly, but not likely to be confused.  1\farcm0 ESE from \\
 &UCM 1300+2907&faint LSB with no published redshift in main beam.  \\

 187&CG 965& Possibly, but not likely to be confused.  2\farcm9 NNW from bright galaxy ($cz$=7941 km s$^{-1}$) on the edge of the main beam.  1/farcm6 W from faint galaxy with no published redshift. \\

 191&\nodata&Noisy baseline due to solar interference.\\
 192&\nodata&4\farcm0 ESE from faint galaxy at first null.  Unlikely to be confused.\\

 193&\nodata&Possibly, but not likely to be confused. 1\farcm3 SSE from faint galaxy in main beam with no published redshift.  4\farcm6 SW from early-type spiral ($cz$ = 7164 km s$^{-1}$) near first null.\\

 194&\nodata&Possibly, but not likely to be confused.  2\farcm7 N from galaxy with no published redshift in outer part of main beam.\\

 205&CG 976&6\farcm1 SSW from galaxy in sidelobe.  Unlikely to be confused.\\

 215&CG 979&Non-detection, two features present in spectrum, but at wrong velocities.\\

 217&CG 980, & \nodata\\
 &UCM 1306+2938&\\

 238&\nodata&Possibly, but not likely to be confused.  1\arcmin S. away from faint LSB galaxy with no published redshift.\\

 252&\nodata&4\farcm2 SW from bright galaxy ($cz$ = 10468 km s$^{-1}$) at first null.  Unlikely to be confused.\\
 256&\nodata& Confused.  2\farcm0 NNW from KISSR 257 ($cz$ = 10,450 km s$^{-1}$)\\
 257&\nodata& Confused.  2\farcm0 SSE from KISSR 256 ($cz$ = 10,501 km s$^{-1}$)\\
 265&\nodata& Confused.  1\farcm7 W from KISSR 266 ($cz$ = 9488 km s$^{-1}$) and 1\farcm7 SW from second galaxy ($cz$ = 9533 km s$^{-1}$).  Both galaxies on edge of main beam.\\

 272&\nodata&Possibly, but not likely to be confused.  0\farcm6 SSW from faint galaxy and 1\farcm1 NW from second faint galaxy.  Both neighbors in main beam, and have no published redshifts.\\  

 280&\nodata&Possibly, but not likely to be confused.  1\farcm2 SE from faint galaxy and 1\farcm6 ENE from second faint galaxy.  Both neighbors in main beam, and have no published redshifts.\\

 286&UCM 1324+2926,&\nodata\\
 &Was 70&\\
\\\\\\\\\\\\\\\\\\\\\\\\\\\\\\\\\\\\\\\\\\\\\\\\\\\\\\\\\\\\\\\\\\\\\\\\\\\\\\\\\\\\\\\\\\\\\\\\\\\\\
 299&\nodata&Possibly, but not likely to be confused.  0\farcm7 E from faint galaxy and 1\farcm5 S from second brighter galaxy.  Both neighbors in main beam, and have no published redshifts.\\
 305&\nodata&   Noisy baseline due to solar interference.\\
 310&UCM 1331+2900,&\nodata\\
 &Was 74&\\

 314&UGC 8578&\nodata\\

 356&\nodata& Confused, but flux revised based on shape of composite profile. 0\farcm6E from highly inclined spiral with no published velocity.  See $\S$ 3.3.\\

 396&Was 81&Possibly, but not likely to be contaminated.  4\farcm5 NE from bright face-on spiral near first null ($cz$=2368 km s$^{-1}$), but observed narrow profile centered at $cz$=2258 km s$^{-1}$.  Weak emission bump at 2368 km s$^{-1}$ not included in measured flux. \\

 398&\nodata& Non-detection.\\

 401&\nodata& Confused.  0\farcm6 SE from KISSR 399 ($cz$=11,245 km s$^{-1}$) and 0\farcm2 SE from KISSR 400 ($cz$=11062 km s$^{-1}$).  Signal probably dominated by KISSR 399 and 401.  Observed quantities for this region are $v_{helio}$=11056 \kms, $\Delta v_{50}$=321 \kms, $\Delta v_{20}$=388 \kms, F.I. = 1.7900 Jy \kms, RMS = 0.8498 mJy, Order = 3, Time = 1146 s, SNR = 46.7.\\

 404&\nodata& Confused.  24\arcsec from KISSR 405 ($cz_{opt}$=9491 km s$^{-1}$). \\

 405&\nodata& Confused.  24\arcsec from KISSR 404 ($cz_{opt}$=9425 km s$^{-1}$).\\

 407&\nodata& 1\farcm9 N from galaxy just outside main beam.  Unlikely to be confused.\\

 460&\nodata& Non-detection.  Early type optical spectrum.  Noisy baseline due to solar interference.\\

 471&\nodata& Possibly, but not likely to be confused.  1\farcm5 SSE from galaxy in main beam with no published redshift.\\

 507&\nodata& Possibly, but not likely to be confused.  3\farcm1 E from edge-on galaxy near first-null with no published redshift.\\

 508&\nodata& 3\farcm7 NNW from galaxy near first null with no published redshift.  Unlikely to be confused.\\

 515&\nodata& Possibly, but not likely to be confused.  1\farcm0 NNE from galaxy in main beam with no published redshift.\\

 528&\nodata&Possibly, but not likely to be confused.  0\farcm8 E from galaxy and 1\farcm0 E from second galaxy.  Both potential companions are in main beam and have no published redshift.\\

 541&\nodata&Possibly, but not likely to be confused.  0.6\farcm SSE from galaxy and 1\farcm2 SW from second galaxy.  Both potential companions are in main beam and have no published redshifts.\\

 561&CG 1230&Possibly, but not likely to be confused.  1\farcm5 NEfrom galaxy in outer part of main beam with no published redshift.\\

 572&\nodata&Possibly, but not likely to be confused.  1\farcm4 W from faint galaxy in main beam with no published redshift.\\

 590&\nodata& Non-detection.\\

 666&\nodata&Possibly, but not likely to be confused.  0\farcm6 NE from galaxy in main beam with no published redshift.\\
\\\\\\\\\\\\\\\\\\\\\\\\\\\\\\\\\\\\\\\\\\\\\\\\\\\\\\\\\\\\\\\\\\\\\\\\\\\\\\\\\\\\\\\\\\\\\\\\\\\\\\\
 675&\nodata& Emission region is HII region in LSB dwarf.\\
 
 742&\nodata& Non-detection.\\
 756&\nodata& Non-detection.\\
 
 785&\nodata&Possibly, but not likely to be confused.  1\farcm0 E from galaxy and 2\farcm9 E from second galaxy.  Both potential companions are in main beam and have no published redshifts.\\\\

 830&\nodata&Possibly, but not likely to be confused.  1\farcm5 W from faint galaxy in main beam with no published redshift.\\
 856&\nodata& Non-detection.  Noisy baseline due to solar interference.\\
 
 999&\nodata& Non-detection.\\
1013&\nodata&Confused, but flux revised based on shape of composite profile.  0\farcm4 ENE from LSB galaxy with no published redshift.  See $\S$ 3.3.\\

1021&\nodata&Possibly, but not likely to be confused.  1\farcm6 ESE from faint galaxy in main beam with no published redshift.\\

1048&UGC 10445&Large apparent size compared to beam. See $\S$ 3.4.\\

\enddata

\end{deluxetable}


\begin{deluxetable}{crrrrc}
\tablewidth{0pt}
\tablecaption{Beam Corrected Fluxes}
\tablehead{
\colhead{KISSR ID}       & \colhead{a$_{25}$}&
\colhead{b$_{25}$}        & \colhead{F.I.$_{obs}$}&
\colhead{$f_c$}          & \colhead{F.I.$_{c}$}   
\\
& \colhead{\footnotesize{(\arcsec)}}& \colhead{\footnotesize{(\arcsec)}} & \colhead{\footnotesize{(Jy$\cdot$km s$^{-1}$)}}&
& \colhead{\footnotesize{(Jy$\cdot$km s$^{-1}$)}} \\  

}

\startdata
1&	27.8	&16.2&	0.812& 1.05& 0.855\\
57&	24.4	& 8.9&	0.755& 1.03& 0.780\\
59&	34.9	&11.3&	1.109& 1.07& 1.184\\
73&	59.9	&53.5&	7.401& 1.33& 9.833\\	
170&	29.9	&7.5 &	0.174& 1.05& 0.183\\
195&	55.4	&15.4&  0.666& 1.16& 0.773\\
207&	36.1	&12.2&	0.865& 1.07& 0.928\\
236&	34.0	&12.2&	0.600& 1.07& 0.639\\
314&	56.8	&23.4&	2.380& 1.18& 2.819\\
460&    28.1    & 9.8&    ND & 1.04& -\\
541&	32.3	&12.8&	0.406& 1.06& 0.431\\
544&	29.8	& 8.8&	0.528& 1.05& 0.553\\
675&	48.1	&32.3&	1.362& 1.17& 1.593\\	
1048$^1$&	168.0	&102.0	&	17.76& 1.78& 31.61\\
\hline
\multicolumn{6}{l}{\tiny $^1$ Listed correction factors computed with double Gaussian HI model of Hewitt et al. 1993.  See $\S$ 3.4.}\\ 	
\hline
\enddata
\end{deluxetable}

\clearpage
\renewcommand{\arraystretch}{0.9}
\begin{deluxetable}{rrrccrr}
\tabletypesize{\small}
\tablecolumns{7}
\tablewidth{0pt}
\tablecaption{KISS HI Sample: Derived Quantities\label{table:tab4}}

\tablehead{

\colhead{KISSR}& 
\colhead{M$_{HI}$}&
\colhead{M$_{HI}$/L$_B$}& 
\colhead{SFR($z_{\odot}$)}&
\colhead{SFR($z$)}& 
\colhead{$\tau(z_{\odot})$}&
\colhead{$\tau(z)$}  
\\

\colhead{ID}& 
\colhead{\tiny{[10$^8$ M$_\odot$]}}& 
\colhead{\tiny{[M$_\odot$/L$_\odot$]}}& 
\colhead{\tiny{[LOG(M$_\odot$/yr)]}}&
\colhead{\tiny{[LOG(M$_\odot$/yr)]}}&
\colhead{\tiny{[LOG(yrs)]}}&
\colhead{\tiny{[LOG(yrs)]}}
\\[0.5ex]  

\colhead{(1)}& 
\colhead{(2)}&
\colhead{(3)}& 
\colhead{(4)}&
\colhead{(5)}& 
\colhead{(6)}&
\colhead{(7)}\\  
}

\startdata  
   1&\phm{<}18.98&\phm{<} 1.01&-0.99&-1.30&\phm{<}10.27&\phm{<}10.58\\[0.0ex]
  32&\phm{<} 4.83&\phm{<} 0.73&-0.93&-1.27&\phm{<} 9.61&\phm{<} 9.96\\[0.0ex]
  40&\phm{<} 7.46&\phm{<} 0.37&-0.64&-1.04&\phm{<} 9.51&\phm{<} 9.91\\[0.0ex]
  49&\phm{<} 8.97&\phm{<} 3.35&-0.47&-0.89&\phm{<} 9.42&\phm{<} 9.84\\[0.0ex]
  52&\phm{<}14.42&\phm{<} 0.60&-0.08&-0.41&\phm{<} 9.24&\phm{<} 9.57\\[0.0ex]
  55&\phm{<} 2.23&\phm{<} 0.63&-1.02&-1.60&\phm{<} 9.37&\phm{<} 9.95\\[0.0ex]
  57&\phm{<}13.94&\phm{<} 1.11&-0.79&-1.11&\phm{<} 9.93&\phm{<}10.25\\[0.0ex]
  59&\phm{<}27.38&\phm{<} 1.44&-1.09&-1.44&\phm{<}10.53&\phm{<}10.88\\[0.0ex]
  61&\phm{<} 1.43&\phm{<} 0.53&-1.38&-1.85&\phm{<} 9.54&\phm{<}10.01\\[0.0ex]
  73&\phm{<} 1.14&\phm{<} 2.56&-2.99&-3.54&\phm{<}11.05&\phm{<}11.60\\[1.5ex]
  75&\phm{<}17.17&\phm{<} 1.50&-1.04&-1.43&\phm{<}10.28&\phm{<}10.67\\[0.0ex]
  85&\phm{<} 0.60&\phm{<} 0.38&-1.30&-1.88&\phm{<} 9.08&\phm{<} 9.66\\[0.0ex]
  91&\phm{<} 8.49&\phm{<} 0.78&-0.69&-1.01&\phm{<} 9.61&\phm{<} 9.94\\[0.0ex]
  96&\phm{<}17.14&\phm{<} 2.10&-0.62&-1.13&\phm{<} 9.85&\phm{<}10.36\\[0.0ex]
  97&\phm{<}26.32&\phm{<} 2.68&-1.10&-1.60&\phm{<}10.52&\phm{<}11.03\\[0.0ex]
 105&\phm{<} 2.20&\phm{<} 2.10&-1.55&-2.10&\phm{<} 9.90&\phm{<}10.44\\[0.0ex]
 108&\phm{<} 2.69&\phm{<} 0.19&-0.62&-1.03&\phm{<} 9.05&\phm{<} 9.46\\[0.0ex]
 109&\phm{<} 4.11&\phm{<} 0.15&-0.52&-0.82&\phm{<} 9.13&\phm{<} 9.43\\[0.0ex]
 115&\phm{<}10.53&\phm{<} 0.52&-0.70&-1.03&\phm{<} 9.72&\phm{<}10.05\\[0.0ex]
 116&\phm{<} 4.96&\phm{<} 0.43&-0.41&-0.85&\phm{<} 9.11&\phm{<} 9.54\\[1.5ex]
 119&\phm{<} 6.90&\phm{<} 1.29&-0.81&-1.31&\phm{<} 9.65&\phm{<}10.15\\[0.0ex]
 120&$<$ 9.82&$<$ 7.80&-1.20&-1.72&$<$10.20&$<$10.71\\[0.0ex]
 125&\phm{<}15.03&\phm{<} 0.78&-0.81&-1.13&\phm{<} 9.98&\phm{<}10.30\\[0.0ex]
 133&\phm{<} 0.88&\phm{<} 0.09&-0.91&-1.22&\phm{<} 8.85&\phm{<} 9.16\\[0.0ex]
 142&$<$17.43&$<$ 0.90&-1.18&-1.53&$<$10.42&$<$10.77\\[0.0ex]
 146$^1$&$<$ 2.24&$<$ 0.69&-1.44&-1.75&$<$10.27&$<$10.58\\[0.0ex]
 148$^1$&$<$ 2.24 &$<$ 0.93&-1.77&-2.14&$<$10.60&$<$10.97\\[0.0ex]
 156&\phm{<} 8.81&\phm{<} 0.95&-1.13&-1.63&\phm{<}10.08&\phm{<}10.57\\[0.0ex]
 170&\phm{<} 3.65&\phm{<} 0.32&-0.74&-1.01&\phm{<} 9.30&\phm{<} 9.57\\[0.0ex]
 171&$<$10.77&$<$ 1.12&-1.21&-1.44&$<$10.24&$<$10.47\\[1.5ex]
 182&\phm{<}10.21&\phm{<} 0.59&-0.65&-1.04&\phm{<} 9.66&\phm{<}10.05\\[0.0ex]
 187&\phm{<} 3.46&\phm{<} 0.71&-1.08&-1.51&\phm{<} 9.62&\phm{<}10.05\\[0.0ex]
 191&\phm{<}17.45&\phm{<} 0.75&-1.01&-1.48&\phm{<}10.25&\phm{<}10.72\\[0.0ex]
 192&\phm{<} 3.37&\phm{<} 0.99&-0.45&-0.83&\phm{<} 8.98&\phm{<} 9.36\\[0.0ex]
 193&\phm{<} 2.28&\phm{<} 0.58&-1.04&-1.45&\phm{<} 9.40&\phm{<} 9.80\\[0.0ex]
 194&\phm{<} 4.44&\phm{<} 0.35&-0.71&-1.07&\phm{<} 9.36&\phm{<} 9.72\\[0.0ex]
 195&\phm{<} 8.94&\phm{<} 0.38&-0.57&-0.90&\phm{<} 9.53&\phm{<} 9.85\\[0.0ex]
 205&\phm{<} 4.35&\phm{<} 0.41&-0.92&-1.41&\phm{<} 9.56&\phm{<}10.05\\[0.0ex]
 207&\phm{<}20.60&\phm{<} 0.99& 0.40&-0.09&\phm{<} 8.91&\phm{<} 9.40\\[0.0ex]
 210&\phm{<} 7.41&\phm{<} 1.13&-1.09&-1.56&\phm{<} 9.96&\phm{<}10.43\\[1.5ex]
 215&$<$13.70&$<$ 0.66&-0.34&-0.68&$<$ 9.48&$<$ 9.81\\[0.0ex]
 217&\phm{<} 8.70&\phm{<} 0.51&-0.72&-1.03&\phm{<} 9.66&\phm{<} 9.97\\[0.0ex]
 236&\phm{<} 9.41&\phm{<} 0.69&-1.17&-1.54&\phm{<}10.14&\phm{<}10.52\\[0.0ex]
 238&\phm{<} 3.91&\phm{<} 0.31&-0.25&-0.63&\phm{<} 8.84&\phm{<} 9.22\\[0.0ex]
 245&\phm{<} 2.59&\phm{<} 0.49&-1.20&-1.55&\phm{<} 9.62&\phm{<} 9.97\\[0.0ex]
 252&\phm{<}17.60&\phm{<} 0.87&-0.70&-1.00&\phm{<} 9.94&\phm{<}10.24\\[0.0ex]
 256&$<$11.86&$<$ 1.28&-0.44&-0.79&$<$ 9.52&$<$ 9.87\\[0.0ex]
 257&$<$ 7.26&$<$ 0.53&-0.46&-0.80&$<$ 9.32&$<$ 9.66\\[0.0ex]
 265&$<$20.33&$<$ 2.21&-0.78&-1.09&$<$10.08&$<$10.40\\[0.0ex]
 271&\phm{<} 6.08&\phm{<} 0.45&-0.63&-1.00&\phm{<} 9.41&\phm{<} 9.78\\[1.5ex]
 272&\phm{<}30.70&\phm{<} 2.62&-1.17&-1.72&\phm{<}10.66&\phm{<}11.21\\[0.0ex]
 278&\phm{<} 1.14&\phm{<} 0.47&-1.86&-2.37&\phm{<} 9.92&\phm{<}10.42\\[0.0ex]
 280&\phm{<} 5.76&\phm{<} 1.62&-1.61&-2.09&\phm{<}10.37&\phm{<}10.85\\[0.0ex]
 286&\phm{<} 5.94&\phm{<} 0.98&-0.31&-0.80&\phm{<} 9.09&\phm{<} 9.57\\[0.0ex]
 299&\phm{<} 2.66&\phm{<} 0.24&-0.71&-1.11&\phm{<} 9.13&\phm{<} 9.53\\[0.0ex]
 310&\phm{<} 4.57&\phm{<} 0.49&-0.13&-0.80&\phm{<} 8.79&\phm{<} 9.46\\[0.0ex]
 305&\phm{<}37.08&\phm{<} 1.55&-0.26&-0.19&\phm{<} 9.83&\phm{<} 9.75\\[0.0ex]
 311&\phm{<} 2.78&\phm{<} 0.38&-0.61&-1.21&\phm{<} 9.05&\phm{<} 9.66\\[0.0ex]
 314&\phm{<} 0.96&\phm{<} 0.53&-1.90&-2.39&\phm{<} 9.88&\phm{<}10.37\\[0.0ex]
 356&\phm{<} 4.80&\phm{<} 0.21&-0.95&-1.28&\phm{<} 9.64&\phm{<} 9.97\\[1.5ex]
 368&\phm{<}10.25&\phm{<} 0.56&-0.54&-0.68&\phm{<} 9.56&\phm{<} 9.69\\[0.0ex]
 386&\phm{<} 6.23&\phm{<} 0.36&-0.78&-1.14&\phm{<} 9.58&\phm{<} 9.94\\[0.0ex]
 396&\phm{<} 1.09&\phm{<} 0.73&-1.53&-2.08&\phm{<} 9.57&\phm{<}10.12\\[0.0ex]
 398&$<$20.35&$<$ 2.23&-0.78&-0.99&$<$10.09&$<$10.30\\[0.0ex]
 401$^2$&$<$29.61&$<$ 2.01&-0.61&-0.97&$<$10.56&$<$10.91\\[0.0ex]
 404&$<$16.36&$<$ 3.55&-1.08&-1.62&$<$10.29&$<$10.84\\[0.0ex]
 405&$<$17.87&$<$ 1.21&-0.77&-1.32&$<$10.02&$<$10.57\\[0.0ex]
 407&\phm{<}27.66&\phm{<} 1.55&-1.13&-1.48&\phm{<}10.57&\phm{<}10.92\\[0.0ex]
 460&$<$18.62&$<$ 0.86&-0.96&-1.29&$<$10.23&$<$10.56\\[0.0ex]
 465&\phm{<} 9.55&\phm{<} 0.84&-1.10&-1.49&\phm{<}10.08&\phm{<}10.47\\[1.5ex]
 471&\phm{<}20.69&\phm{<} 2.51&-1.08&-1.76&\phm{<}10.40&\phm{<}11.08\\[0.0ex]
 505&\phm{<} 2.06&\phm{<} 0.33&-1.22&-1.74&\phm{<} 9.54&\phm{<}10.06\\[0.0ex]
 507&\phm{<} 3.88&\phm{<} 0.23&-1.08&-1.43&\phm{<} 9.67&\phm{<}10.02\\[0.0ex]
 508&\phm{<} 6.76&\phm{<} 0.73&-0.36&-0.76&\phm{<} 9.19&\phm{<} 9.59\\[0.0ex]
 515&\phm{<}28.01&\phm{<} 1.47&-0.69&-0.95&\phm{<}10.14&\phm{<}10.39\\[0.0ex]
 518&$<$19.36&$<$ 0.79&-0.17&-0.39&$<$ 9.45&$<$ 9.67\\[0.0ex]
 528&\phm{<}12.88&\phm{<} 1.49&-0.66&-1.19&\phm{<} 9.77&\phm{<}10.30\\[0.0ex]
 541&\phm{<} 1.71&\phm{<} 0.35&-0.41&-0.92&\phm{<} 8.65&\phm{<} 9.15\\[0.0ex]
 544&\phm{<}17.87&\phm{<} 0.83&-0.42&-0.77&\phm{<} 9.67&\phm{<}10.02\\[0.0ex]
 561&\phm{<} 5.02&\phm{<} 1.86&-1.43&-1.91&\phm{<}10.13&\phm{<}10.61\\[1.5ex]
 564&\phm{<}14.58&\phm{<} 0.89&-0.70&-1.05&\phm{<} 9.86&\phm{<}10.22\\[0.0ex]
 572&\phm{<} 1.58&\phm{<} 0.39&-1.60&-1.96&\phm{<} 9.80&\phm{<}10.16\\[0.0ex]
 590&$<$14.92&$<$ 2.98&-0.78&-1.13&$<$ 9.95&$<$10.31\\[0.0ex]
 666&\phm{<} 3.89&\phm{<} 1.12&-0.69&-1.37&\phm{<} 9.28&\phm{<} 9.96\\[0.0ex]
 675&\phm{<}18.42&\phm{<} 1.25&-1.40&-1.99&\phm{<}10.67&\phm{<}11.25\\[0.0ex]
 678&\phm{<}12.53&\phm{<} 1.53&-0.57&-0.91&\phm{<} 9.67&\phm{<}10.00\\[0.0ex]
 742&$<$19.63&$<$ 2.32&-0.80&-1.23&$<$10.09&$<$10.52\\[0.0ex]
 756&$<$33.45&$<$ 1.77&-0.46&-0.69&$<$ 9.99&$<$10.21\\[0.0ex]
 757&\phm{<}19.97&\phm{<} 1.14&-0.37&-0.86&\phm{<} 9.67&\phm{<}10.16\\[0.0ex]
 785&\phm{<} 9.68&\phm{<} 1.59&-0.79&-1.35&\phm{<} 9.78&\phm{<}10.34\\[1.5ex]
 803&\phm{<}10.42&\phm{<} 0.81&-0.73&-1.28&\phm{<} 9.75&\phm{<}10.29\\[0.0ex]
 830&\phm{<}19.87&\phm{<} 0.81&-0.44&-0.74&\phm{<} 9.74&\phm{<}10.04\\[0.0ex]
 856&$<$ 5.02&$<$ 5.50&-1.57&-2.24&$<$10.27&$<$10.94\\[0.0ex]
 882&\phm{<} 5.62&\phm{<} 0.45&-0.78&-0.92&\phm{<} 9.53&\phm{<} 9.67\\[0.0ex]
 891&\phm{<} 8.54&\phm{<} 0.56&-0.75&-1.11&\phm{<} 9.68&\phm{<}10.05\\[0.0ex]
 956&\phm{<} 6.62&\phm{<} 1.00&-1.01&-1.38&\phm{<} 9.84&\phm{<}10.20\\[0.0ex]
 979&\phm{<} 8.21&\phm{<} 0.70&-0.89&-1.31&\phm{<} 9.81&\phm{<}10.22\\[0.0ex]
 998&\phm{<} 1.15&\phm{<} 0.15&-0.55&-1.00&\phm{<} 8.61&\phm{<} 9.06\\[0.0ex]
 999&$<$23.68&$<$ 0.98&-0.95&-1.27&$<$10.32&$<$10.65\\[0.0ex]
1005&\phm{<} 4.35&\phm{<} 0.35&-1.38&-1.76&\phm{<}10.02&\phm{<}10.40\\[1.5ex]
1011&\phm{<} 5.60&\phm{<} 1.65&-2.14&-2.68&\phm{<}10.88&\phm{<}11.43\\[0.0ex]
1013&\phm{<}10.31&\phm{<} 0.77&-0.94&-1.45&\phm{<} 9.96&\phm{<}10.46\\[0.0ex]
1014&\phm{<} 5.24&\phm{<} 1.01&-0.79&-1.20&\phm{<} 9.51&\phm{<} 9.92\\[0.0ex]
1021&\phm{<} 1.55&\phm{<} 0.99&-1.76&-2.26&\phm{<} 9.95&\phm{<}10.45\\[0.0ex]
1048&\phm{<}16.98&\phm{<} 2.79&-3.15&-3.52&\phm{<}12.38&\phm{<}12.75\\[0.0ex]
1091&\phm{<} 1.40&\phm{<} 0.25&-0.89&-1.37&\phm{<} 9.04&\phm{<} 9.52\\[0.0ex]
1112&\phm{<}27.91&\phm{<} 3.26&-0.93&-1.37&\phm{<}10.38&\phm{<}10.82\\[0.0ex]
1121&\phm{<} 7.56&\phm{<} 0.63& 0.01&-0.48&\phm{<} 8.86&\phm{<} 9.36\\[0.0ex]
1123&\phm{<} 3.74&\phm{<} 0.53&-0.51&-0.99&\phm{<} 9.08&\phm{<} 9.56\\[0.0ex]
\hline
\multicolumn{7}{l}{\tiny $^1$ 21-cm emission dominated by U8033.  Upper limits on HI mass computed from footnote 1 of Table 1.}\\
\multicolumn{7}{l}{\tiny $^2$ 21-cm emission probably dominated by KISSR 399 and 401.  Upper limits on HI mass computed from}\\[-0.5ex]
\multicolumn{7}{l}{\tiny \phm{$^2$} footnote 2 of Table 1.}\\
\hline

\enddata
\end{deluxetable}

\end{document}